\documentclass[twocolumn,floatfix,superscriptaddress,a4paper,showpacs,showkeys,nofootinbib,notitlepage]{revtex4-1}
\usepackage[colorlinks=true,linktocpage=true,linkcolor=blue,citecolor=blue,allcolors=blue]{hyperref}
\usepackage{epsfig}
\usepackage{latexsym}
\usepackage[utf8]{inputenc}
\usepackage{xspace}
\usepackage{indentfirst}
\usepackage{enumitem}
\usepackage{color}
\usepackage{placeins}
\usepackage{setspace}
\usepackage{lipsum}

\usepackage{hyperref}

\usepackage{amsmath}
\usepackage{amssymb}
\usepackage[english]{babel}
\usepackage{url}
\graphicspath{{figs/}}

\newcommand{\mean}[1]{\langle #1 \rangle}
\newcommand{\eq}[1]{\begin{align} #1 \end{align}}

\newcommand{\be}{\begin{equation}}
\newcommand{\ee}{\end{equation}}

\begin{document}
\title{Phase transition amplification of proton number fluctuations in nuclear collisions\\ from a transport model approach}
\author{Oleh Savchuk}
\affiliation{Facility for Rare Isotope Beams, Michigan State University, East Lansing, MI 48824 USA}
\affiliation{Bogolyubov Institute for Theoretical Physics, Kyiv, Ukraine}
\affiliation{
GSI Helmholtzzentrum f\"ur Schwerionenforschung GmbH, Planckstr. 1, D-64291 Darmstadt, Germany}
\author{Roman V. Poberezhnyuk}
\affiliation{Bogolyubov Institute for Theoretical Physics, Kyiv, Ukraine}
\affiliation{Frankfurt Institute for Advanced Studies, Giersch Science Center, Ruth-Moufang-Str. 1, D-60438 Frankfurt am Main, Germany}
\author{Anton Motornenko}
\affiliation{Frankfurt Institute for Advanced Studies, Giersch Science Center, Ruth-Moufang-Str. 1, D-60438 Frankfurt am Main, Germany}
\author{Jan Steinheimer}
\affiliation{Frankfurt Institute for Advanced Studies, Giersch Science Center, Ruth-Moufang-Str. 1, D-60438 Frankfurt am Main, Germany}	

\author{Mark~I.~Gorenstein}
\affiliation{Bogolyubov Institute for Theoretical Physics, Kyiv, Ukraine}
\affiliation{Frankfurt Institute for Advanced Studies, Giersch Science Center, Ruth-Moufang-Str. 1, D-60438 Frankfurt am Main, Germany}
\author{Volodymyr Vovchenko}
\affiliation{Physics Department, University of Houston, Box 351550, Houston, TX 77204, USA}
\affiliation{Institute for Nuclear Theory, University of Washington, Box 351550, Seattle, WA 98195, USA}
\affiliation{Frankfurt Institute for Advanced Studies, Giersch Science Center, Ruth-Moufang-Str. 1, D-60438 Frankfurt am Main, Germany}

\date{\today}

\begin{abstract}
The time evolution of particle number fluctuations in nuclear collisions at intermediate energies~($E_{\rm lab} = 1.23-10A$~GeV) is studied by means of the UrQMD-3.5 transport model.
The transport description incorporates baryonic interactions through a density-dependent potential. This allows for an implementation of a first order phase transition including a mechanically unstable region at large baryon density.
The scaled variance of the baryon and proton number distributions is calculated in the central cubic spatial volume of the collisions at different times. A significant enhancement of fluctuations associated with the unstable region is observed.
This enhancement persists to late times reflecting a memory effect for the fluctuations.
The presence of the phase transition has a much smaller influence on the observable event-by-event fluctuations of protons in momentum space.
\end{abstract}

\maketitle
\section{Introduction}

The structure of the QCD phase diagram 
is one of the major open questions in high energy physics with great theoretical and experimental efforts on the line~\cite{Busza:2018rrf,Bzdak:2019pkr}. 
Most experimental information for these studies  comes from observables in nucleus-nucleus (A+A) collisions.
The hypothetical first order QCD phase transition at finite baryon density ending in a critical point can be probed in A+A
collisions at intermediate energies.
The dynamical description of A+A collisions 
is based usually on
hydrodynamics and phenomenological transport models~\cite{PhysRevC.72.064901,Petersen_2019,Bass:1998ca,Bleicher:1999xi,Bleicher:2022kcu,Li:2016uvu,Steinheimer:2018rnd,PhysRevLett.75.596}.
Phase transition should show itself in the collapse of the directed flow ~\cite{Rischke:1995pe,Stoecker:2004qu,Brachmann:1999xt,Brachmann:1999mp,Oliinychenko:2022uvy}, fluctuations and correlations~\cite{Stephanov:1998dy,Stephanov:1999zu,Jeon:2000wg,Asakawa:2000wh,Bzdak:2019pkr}, as well as light nuclei enhancement~\cite{Sun:2017xrx}. Another probe is electromagnetic radiation, in particular measurements of dilepton emission, is a promising tool for studying the properties of dense and hot matter~\cite{Shuryak:1978ij,McLerran:1984ay,Bratkovskaya:1996qe,Rapp:1999ej,Savchuk:2022aev}.

During the freeze-out stage of A+A collisions the interactions between particles are expected to be weak. At this time, the multiplicities $\mean{N_i}$ of the $i^{th}$ different measurable hadron species
in A+A collisions 
appear to be
reasonably well described within the framework of an ideal hadron resonance gas 
model~\cite{STAR:2017sal,Harabasz:2020sei,Motornenko:2021nds}. 
Thermal fits are not sensitive to a possible phase transition during the early stages of evolution in A+A reaction, the extracted S/A could be in indirect ways~\cite{Bumnedpan:2022lma}. 
Therefore one should try to study
event-by-event particle number fluctuations \cite{Luo:2017faz,Bzdak:2019pkr}. 
Here, the hope is to understand the equation of state (EoS) properties from measurements of intensive combinations of the central moments of the event-by-event multiplicity distribution of the proton number $N_i$, $\langle(\Delta N_i)^{2}\rangle\equiv \sigma^2$, $\langle (\Delta N_i)^{3}\rangle$, $\langle (\Delta N_i)^{4}\rangle$, etc, where 
$\Delta N_i \equiv N_i -\langle N_i \rangle$. In particular, the scaled variance $\omega$, 
(normalized) skewness $S\sigma$, and  kurtosis $\kappa\sigma^{2}$ of the particle number distribution are defined as follows:
\eq{ \omega[N_i] & =\frac{\sigma^2}{\mean{N_i}}~=~\frac{\kappa_2}{\kappa_1}~
,\label{omega}\\
S\sigma[N_i] & = \frac{\langle(\Delta N_i)^{3}\rangle}{\sigma^{2}}~=~\frac{\kappa_3}{\kappa_2}~, \label{skew}\\
 \kappa  \sigma^{2}[N_i] & = \frac{\langle(\Delta N_i)^{4}\rangle-3\langle(\Delta N_i)^{2}\rangle^2}{\sigma^{2}}~=~\frac{\kappa_4}{\kappa_2}~,\label{kurt}
 }
where $\kappa_n$ are 
the $n$-th cumulants of the $N_i$-distribution.
Such volume-independent (intensive) measures of number fluctuations  (\ref{omega}-\ref{kurt}) 
can also be applied to conserved charges such as net baryon number $B$ and electric charge $Q$.

The amplification of baryon number fluctuations due to the first order phase transition has been predicted in fluid-dynamical simulations of heavy-ion collisions~\cite{Steinheimer:2012gc,Steinheimer:2013glaa}. The equilibration time $\tau$ for fluctuations is typically larger than for mean quantities~\cite{Asakawa:2000wh,Jeon:2000wg}. 
This is especially true for high order fluctuation measures, such as the kurtosis in the vicinity of the critical point~(critical slowing down) \cite{Berdnikov_2000,BLUHM2020122016}.
Thus, one can expect a memory effect for particle number fluctuations during the fast expansion processes: i.e. large fluctuations generated in the intermediate stages of A+A collisions partially survive to the freeze-out stage~\cite{Kitazawa:2012at,Mukherjee:2015swa}. 

Previously, the memory effect was addressed within hydrodynamics with explicit evolution of fluctuations (see, e.g., Ref.~\cite{Pradeep:2022mkf} and references therein) or evolution equations for the cumulants~\cite{Mukherjee:2015swa}.
Significant memory effects on the net proton cumulants
were predicted if the system trajectory passes the vicinity of the critical point at supercritical temperatures. 
The goal of the present paper is to explore the  enhancement of event-by-event fluctuations within a dynamical description of A+A collisions and analyze whether it may survive to the freeze-out stage. The advantage of this approach is that the dynamical evolution of particle number fluctuations, in phase space, is treated consistently throughout the system's evolution. 

The paper is organized as follows. 
In Sec.~\ref{sec-2}
the  UrQMD model with the two types of EoS is formulated. 
In Sec.~\ref{sec-3}  the results for baryon  and proton number fluctuations are presented, both in coordinate and momentum space.
The summary in Sec.~\ref{sec-5} closes the paper.

\noindent \section{\label{sec-2}
UrQMD simulations with a phase transition
}
The UrQMD  transport model 
\cite{Bass:1998ca,Bleicher:1999xi,Bleicher:2022kcu} describes A+A collisions 
in terms of the explicit propagation of hadrons in phase space, their elastic and inelastic two-body reactions, and decay of unstable particles. In its cascade version the effective equation of state of UrQMD resembles a non-interacting hadron resonance gas \cite{Bravina:1998pi,Zabrodin:2009fz}. To include a more realistic EoS and even a phase transition, a density dependent potential is introduced in the QMD part of the simulation. The interactions among baryons are implemented via a
density dependent potential energy per baryon $V(n_B)$. 
This allows to incorporate any density-dependent
EoS in the non-relativistic Hamilton equations of motion (see~Refs.~\cite{OmanaKuttan:2022the,Steinheimer:2022gqb} for more details)\footnote{Note, that the extension of our approach to a relativistic description is not a trivial task due to the fact that QMD is not a local mean field model. A possible solution to this problem has been presented in the RQMD approach \cite{Sorge:1989vt,Nara:2020ztb}. However, the RQMD approach would also require the use of the scalar in addition the vector potential energy density and a method how these can be extracted from any EoS-model would have to be developed. This problem will be addressed in future works.}. 

For the present work we use an EoS which was derived from the Chiral SU(3)-flavor parity-doublet Polyakov-loop quark-hadron mean-field model (CMF) \cite{Steinheimer:2010ib, Steinheimer:2011ea, Mukherjee:2016nhb, Motornenko:2018hjw}, both in its
most recent version~\cite{Motornenko:2019arp},
and in a modified version
that contains an additional first order phase transition
(PT),
denoted further as PT. 
The CMF model incorporates a realistic description of nuclear matter with nuclear incompressibility  of $K_0=267$ MeV, chiral symmetry breaking in the hadronic and quark sectors, as well as an effective deconfinement transition.

In the case of the PT model a simple augmentation is used in order to implement an additional phase transition.
To provide a metastable state 
at large baryon densities 
the original mean-field potential of the CMF model is truncated at density $n_B^{cut} = 2.6 n_0$,
where $n_0=0.16$~fm$^{-3}$ is the nuclear matter saturation density. 
For $n_B>n_B^{cut}$ the potential is then shifted by $\Delta n_B= 2.6n_0$. 
The mean-field energy between $n_B^{cut}<n_B<n_B^{cut}+\Delta n_B$ is interpolated by a third order polynomial in order to create a second minimum in the energy per baryon $V(n_B)$ and to ensure that its derivative is a continuous function. Note, that this procedure modifies the CMF EoS only at high baryon densities, leaving low-density description consistent with nuclear matter properties and lattice QCD constraints. For details see Refs.~\cite{Savchuk:2022aev,Li:2022iil,Steinheimer:2022gqb} where this procedure was introduced.

The specific choice of the value for $n_{cut}$ is motivated by two factors:\\
1. The transition should be reachable with heavy ion collision experiments (which limits the density range to $n_B^{cut} < 4-5 n_0$) \cite{OmanaKuttan:2022the}.\\
2. The density is higher than $2 n_0$ to avoid discrepancies with available constraints from heavy ion collisions and astrophysical observations \cite{Huth:2021bsp}.

\begin{figure}[t]
\includegraphics[width=.49\textwidth]{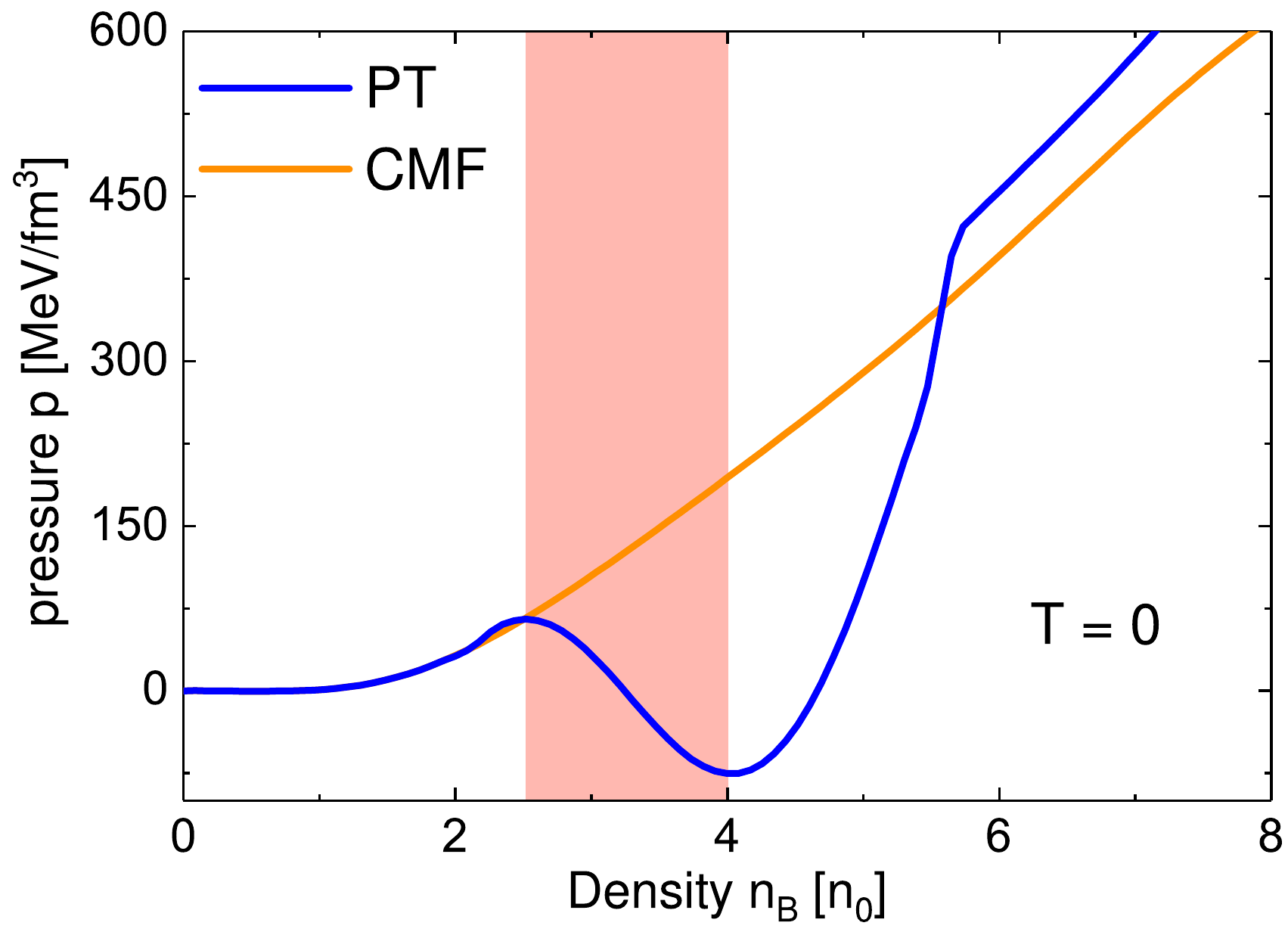}
\caption{
Pressure as a function of density at $T=0$ for the two scenarios under investigation. The CMF EoS which has no phase transition is shown as the orange line and the blue line depicts the construction with a phase transition. The range in density where the pressure gradient is negative is called the spinodal region (indicated by the red shaded area). 
}
\label{fig:press}
\end{figure}

Figure \ref{fig:press} compares the effective pressure of these two equations of state at $T=0$ as a function of density. In this figure also the region of mechanical instabilities is highlighted. If the system created in a collision enters this area, the interactions will drive the system to rapidly separate into the two coexisting phases, a mechanism well known as spinodal decomposition. This phase separation will then lead to the formation of clusters and consequently to an enhancement of the baryon fluctuations in coordinate space. At this point it is important to note that the process of phase separation and the properties (like the surface energy) of the clusters which are created depends strongly on the finite range interactions present in the system \cite{Randrup:2003mu}. In the QMD approach used in this work, the effective range of the interactions is governed by the width of the Gaussian wave packages which are used to calculate the interaction density. While this parameter was fixed to give a proper description of nuclear matter and nuclei it can be argued that for another phase transition (quark-hadron for example) this range may be different. We therefore expect that our results on the properties of the baryon clumps created may not be very precise and only can give a qualitative picture of the dynamics that are expected. Fortunately it has been shown in a previous work, where the role of the finite range interactions on the spinodal clumping in nuclear collisions was studied, it was found that the quantitative effect on the density fluctuation is indeed rather small \cite{Steinheimer:2013glaa}. Thus, the following results may be parameter dependent on a quantitative level, the general results of our work are likely quite robust as observed in previous studies.

\begin{figure}[t]
\includegraphics[width=.49\textwidth]{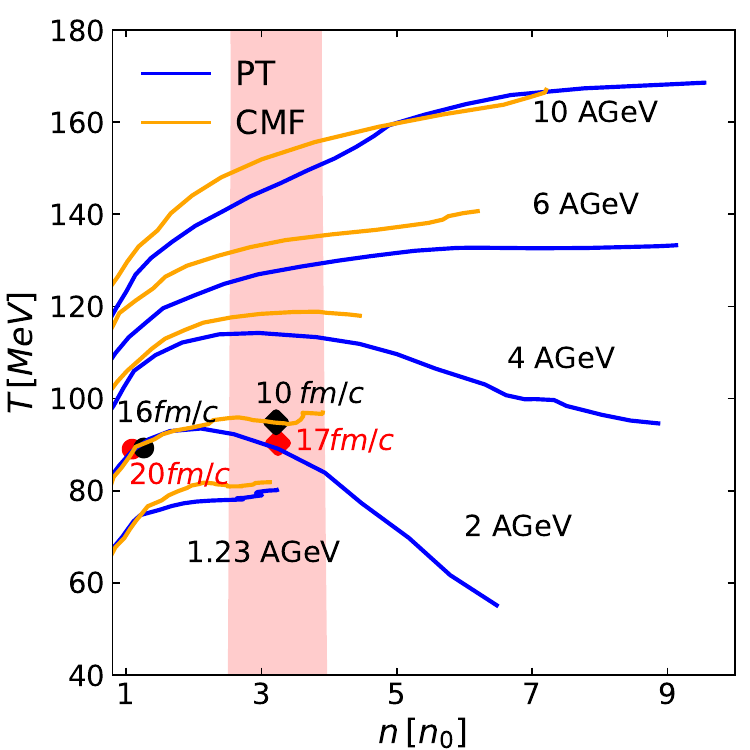}
\caption{
The trajectory of central Au+Au collisions on the phase diagram in the 
plane of baryon density $n_B$ and temperature $T$ at collision energies $E_{\rm lab} = 1.23,\,2,\,4,\,6,$ and $10A$~GeV~(from bottom to top).
The trajectories correspond to the central cubic box of volume $V=27$~fm$^3$.
The blue and orange lines represent the PT and CMF equations of state, respectively.
The diamonds and circles correspond to densities $n_B=3.2~n_0$ and $n_B=1.2~n_0$, respectively.
Red symbols correspond to times of $17fm/c$ and $20fm/c$ and black ones to $10fm/c$ and $16fm/c$.
The red shaded area shows the mechanically unstable region bounded by the spinodals for the PT EoS. 
}
\label{fig:Tn}
\end{figure}

To study the time evolution of the bulk matter as well as the fluctuations caused by the phase transition, we simulate $50000$ collision events for very central ($b<2$ fm or $2\%$ most central) Au-Au collisions in the SIS18/SIS100 and RHIC-BES energy range of $E_{\rm lab}=1.23\div 10A$~GeV.

In particular, we study the time dependence of the net baryon, $B\equiv N_B-N_{\bar{B}}$, and net proton, ${\rm p}\equiv N_p - N_{\bar{p}}$, numbers inside 
a cubic volume of size $V=27$~fm$^3$ located in the geometrical center in the center of mass frame of Au+Au reaction.
Note, that in the considered region of collision energies $E_{\rm lab}$ below 10~AGeV one has $N_{\bar{B}}\ll N_B$, thus the antibaryons and antiprotons play essentially no role.

The size of the central volume is not unique and we have checked that moderate changes of the box size will only modify the extracted second order cumulants according to the fraction of the baryon number enclosed in the box $\alpha$, i.e. by the factor $1-\alpha$ (see Refs.~\cite{Vovchenko:2020tsr,Poberezhnyuk:2020ayn,Vovchenko:2020gne}) and therefore our conclusions remain independent on the box size. On the other hand a too large volume will suppress the fluctuations due to the conservation of the baryon number and a too small volume will simply give the Poissonian baseline result. As the range of the QMD interaction is determined as a few fm (given by the range parameter $L$ in the Gaussian wave package) it is reasonable to choose a box length which is as big or larger than the interaction range but as small as possible to avoid effects from conservation, thus a box length of 3~fm was selected.

Figure~\ref{fig:Tn} presents the UrQMD results for the event averaged expansion trajectories  of the central cubic volume in the $(n_B,T)$-plane for different beam energies.
These trajectories are started at the maximal $n_B$ values achieved at given $E_{\rm lab}$. At this stage the momentum spectra of baryons inside the central volume $V$ are consistent with the thermal Maxwell-Boltzmann distribution.
The temperature of the central volume is extracted from particle and energy densities in the cell, matching to the CMF-EoS.
The trajectories end at $n_B=n_0$. At this stage of the expansion one still has $N_B=n_0V \cong 4$ baryons inside the central volume. 

In  case of the PT model, the effects of the softening of the EoS
occur in the spinodal region.  
This results in a significant increase of the compression and large baryon densities. 

Also, the time evolution of the density is strongly affected by the EoS and the times at which the system, created at $E_{\rm lab}= 2A$ GeV, reaches $1.2$ and $3.2$ nuclear saturation densities is pointed out with the red and black symbols.
During its evolution the created system with a PT evolves through mixed phase region
which leads to a significantly longer expansion time and possibly to an increase of the fluctuations in coordinate space.
The purpose of this work is to test to which extent these large fluctuations in the coordinate space 
may survive to the late stages of the collisions
and whether they can be observed in a momentum space acceptance similar to experimental detection.

\begin{figure*}
\includegraphics[width=.98\textwidth]{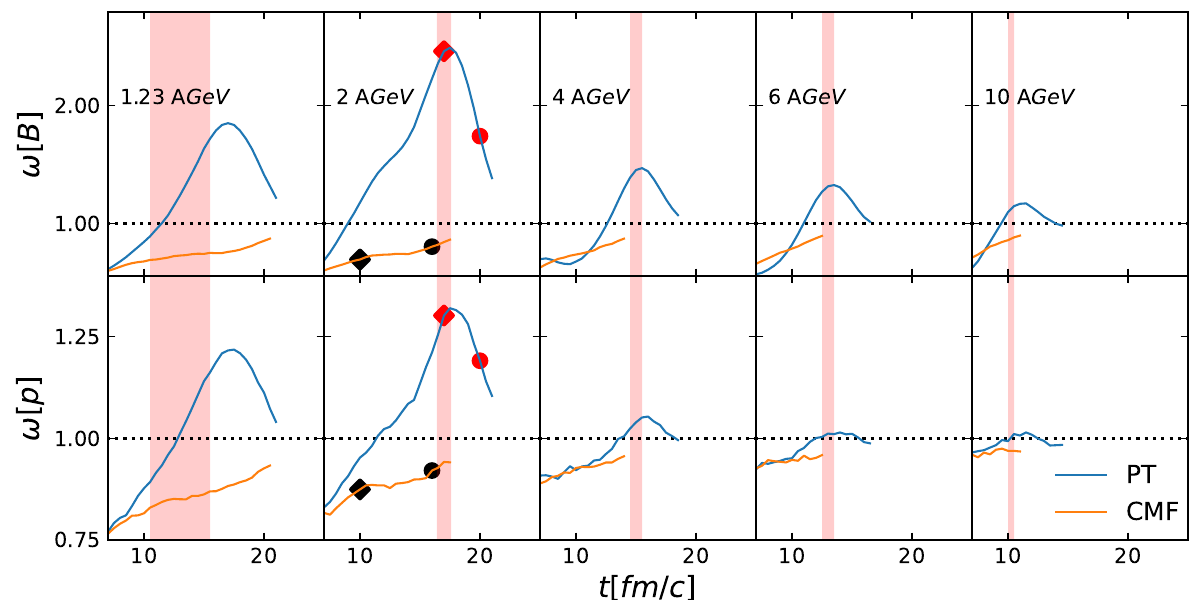}
\caption{\label{var-coordinate}The UrQMD results in central Au+Au collisions for the scaled variance of net baryon (upper panel) and net  proton (lower panel) distributions inside a central cubic box of volume $V=27fm^3$ as a function of the collision time in the center of mass system. 
The blue and orange lines correspond to PT and CMF EoS, respectively. Dotted horizontal lines show the Poisson baseline of $\omega = 1$.
The time $t=0$ is taken as the moment of maximal baryon density in the central box. The lines are stopped at $t$ values that correspond to $n_B=n_0$ in the central cell.
The symbols for $E_{\rm lab} = 2A$~GeV have the same meaning as in Fig.~\ref{fig:Tn}.
For each collision energy the unstable spinodal region is shown by the shaded area.
}
\end{figure*}

\section{Event-by-Event  Fluctuations}\label{sec-3}

Baryon number fluctuations are sensitive probes of the interactions \cite{Li:2016uvu,Steinheimer:2018rnd}.
However, several additional factors affect these fluctuations, including baryon 
and electric charge conservation, non-equilibrium dynamics, resonance formation and decay, as well as finite size effects. The detailed microscopical UrQMD simulations naturally include these effects.

\subsection{Coordinate space}

Figure~\ref{var-coordinate} (upper panel)  presents the UrQMD results for the scaled variance  of baryon number fluctuations, $\omega[B]$, in the central volume $V=3^3$~fm$^3$ as functions of time for different collision energies\footnote{To evaluate the time dependence, we add unformed baryons from string decays to the proton number. This is a reasonable assumption as the string cross section is still very small in the beam energy range considered and we are not studying the effects of a change in degrees of freedom.}. The results are compared to the Poisson distribution baseline, $\omega=1$, which corresponds to the production of uncorrelated particles and usually is approached late when essentially no particles are present in the central volume.
Deviations from the Poisson baseline are evident at intermediate times.
In the CMF case one has $\omega<1$ which is driven by the global charge conservation effects.
However, in the case of PT, 
one observes 
$\omega > 1$, and the maximum values are reached when the system crosses the PT region.
This behavior of fluctuations  
reflects the enhancement of fluctuations due to the instabilities associated with the first order PT.

The two EoS are indistinguishable at lower baryon densities, in particular at $n = n_0$, reached at later stages of the collision.
The scaled variance of baryon number distribution at time moments corresponding to $n = n_0$ inside the box is still different in the two scenarios despite the fact that the two EoS
become identical at this stage.
We find that the scaled variance is enhanced by 
a factor of 2
at $E_{\rm lab}=2A\,$GeV at $2n_0$ in the PT scenario.
Thus, in coordinate space one observes a memory effect for the fluctuations in the fast expanding system.

\begin{figure*}[t!]
\includegraphics[width=.98\textwidth]{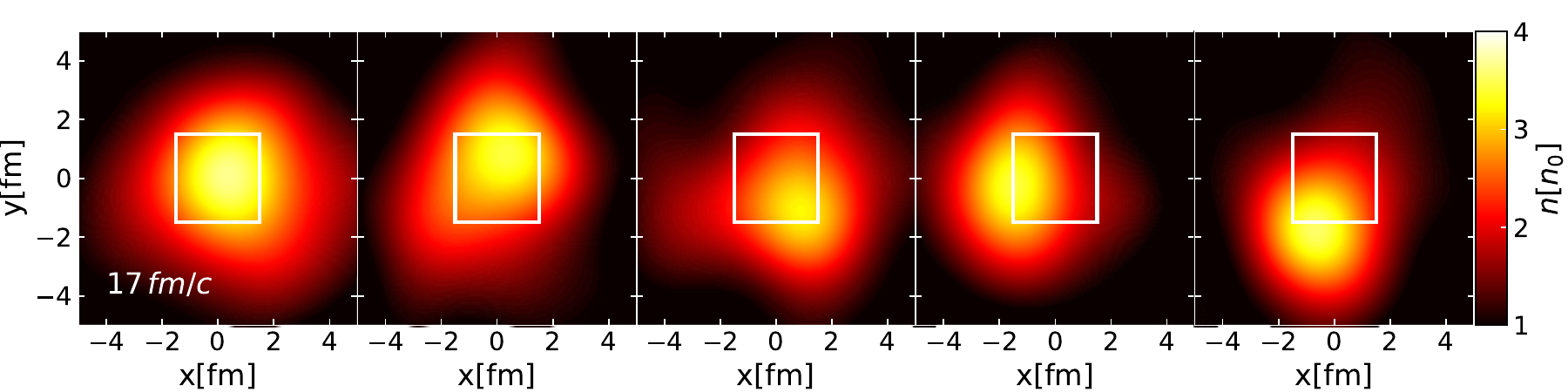}
\caption{\label{dens-coord} Baryon number densities in five different events in the transverse coordinate plane $z=0$ at $t=17~$fm/$c$ in Au+Au central collisions at $E_{\rm lab}=2A$\,GeV for the PT EoS. The time moment of $t = 17$~fm/$c$ corresponds to average $n_B=3.2\,n_0$ inside the central box when the matter is in the mixed liquid-gas state. The projections of the central box on the $z=0$ plane are shown by the white squares.  
}
\end{figure*}

The enhancement of $\omega[B]$ 
due to the presence of the PT is most prominent 
at $E_{\rm lab}=2A\,$GeV. 
This enhancement decreases  rapidly  with further increase of collision energy. 
This is because the system
spends less time inside the PT region during the expansion at the higher collision energies. The shorter time intervals  become  insufficient to generate large fluctuations.

In the experiment, it is problematic to measure all baryons due to difficulties with the detection of neutral particles such as neutrons.
For this reason, fluctuations of the proton number are commonly used as a proxy for the baryon number.
The lower panel of Fig.~\ref{var-coordinate} depicts the UrQMD results for the scaled variance of the proton number distribution, $\omega[{\rm p}]$.
This quantity retains the qualitative features observed for $\omega[B]$ but shows smaller deviations from the Poisson baseline of unity.
This can be understood as an additional acceptance effect modeled by a binomial distribution \cite{Kitazawa:2012at,Bzdak:2012ab,Savchuk:2019xfg}.
We find that $\omega[{\rm p}]$ is a good qualitative proxy of $\omega[B]$, while quantitatively the fluctuation signals are suppressed by about 50\%.
Note, that at early times a significant part of baryons is present as resonance states which are not included in the proton number, reducing the fraction of protons among all baryons.

\begin{figure*}[t!]
\includegraphics[width=.49\textwidth]{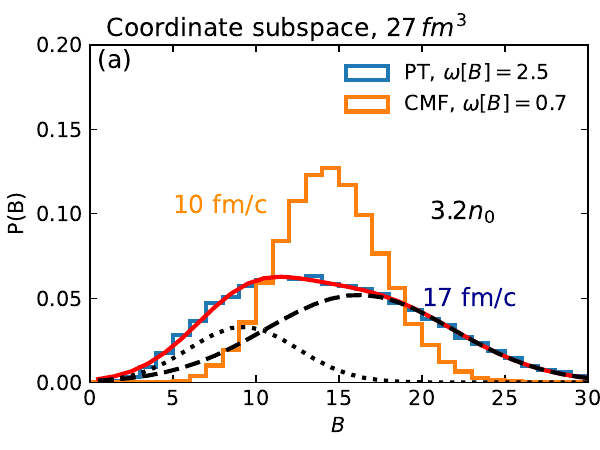}
\includegraphics[width=.49\textwidth]{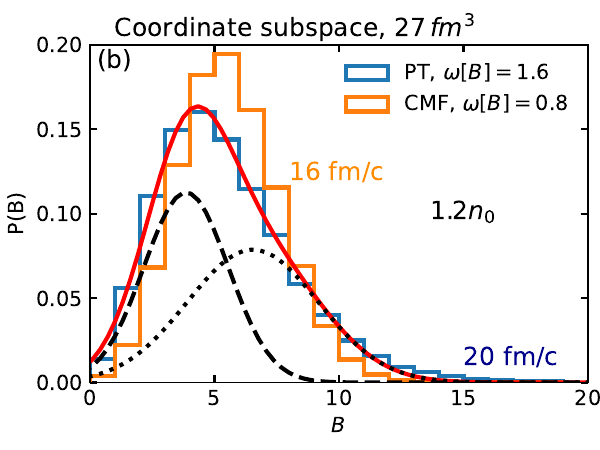}
\includegraphics[width=.49\textwidth]{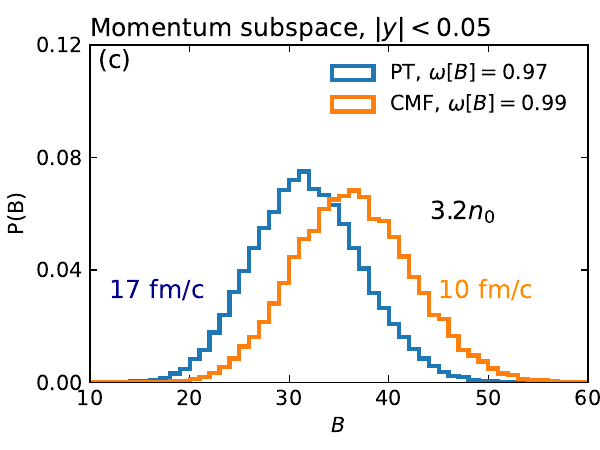}
\includegraphics[width=.49\textwidth]{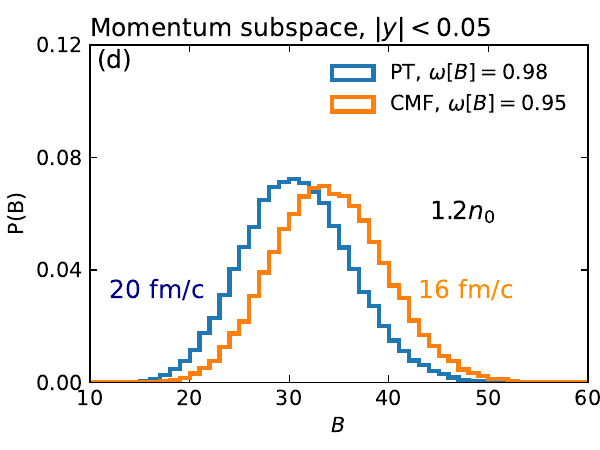}
\caption{\label{distr} Baryon number distributions $P(B)$ inside the central box of volume $V=27~$fm$^3$ (upper panels) and inside $|y|< 0.05$ rapidity interval (lower) panels. Left panels correspond to the dense $n_B=3.2n_0$ system that for PT EoS is inside the unstable region bounded by spinodals. Right panels correspond to the systems at 
$n_B=1.2n_0$ where both CMF and PT EoS have the same properties.
The fits of $P(B)$ for the PT EoS with two Gaussian distributions -- dotted and dashed lines -- are shown in the upper panels by the red solid lines. 
}
\end{figure*}

\begin{figure}[t!]
\includegraphics[width=.49\textwidth]{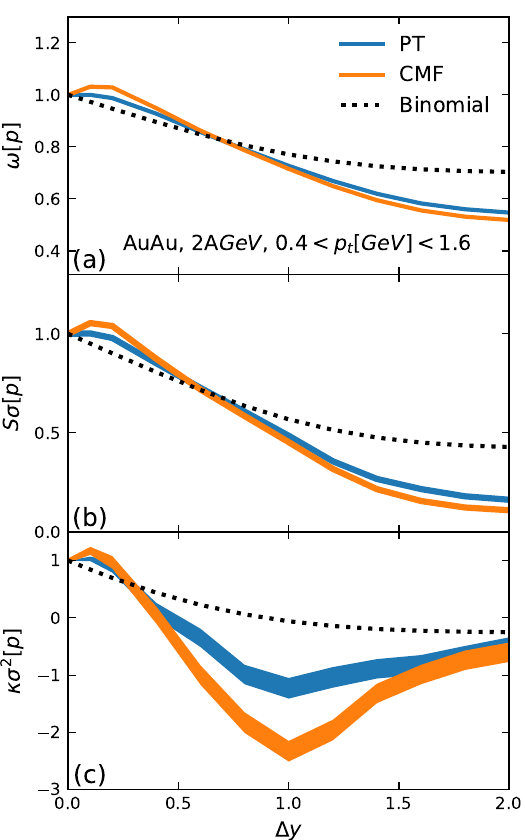}
\caption{\label{protons-final} 
The UrQMD results for the scaled variance $\omega[{\rm p}]$, skewness $S\sigma[{\rm p}]$, and kurtosis $\kappa \sigma^2[{\rm p}]$
of proton number distribution as functions of the rapidity acceptance interval $\Delta y$
for Au-Au collisions at $E_{lab}=2A\,$GeV.
}
\end{figure}

To understand the enhancement of fluctuations due to the PT let us consider its possible dynamics during an expansion.
Two different scenarios are possible.
In case of a slow change in thermodynamic properties, i.e., if the system is constantly in equilibrium, one expects to see nucleation and growth of the new phase. 
However, nucleation is usually suppressed by a potential barrier as a function of spatial coordinate that needs to be crossed in order to achieve a stable droplet of the new phase. This process generally requires a long time which exceeds the time scales reached in heavy-ion collisions. 
Another possibility is spinodal decomposition. When a system is inside the spinodal region on the phase diagram, the associated mechanical instability makes fast separation into phases possible. This separation creates regions of high and low density that correspond to two stable states of matter and, on even-by-event basis, can create high fluctuations in the coordinate space.
The central box of volume $V=27\,$fm$^3$ in the coordinate space can, in each event, contain either the gas or the liquid phase, leading to two peaks in the particle number distribution as separate contributions from 
the two phases~\cite{Sorensen:2020ygf}.

To point out this effect, figure \ref{dens-coord} shows five different, single event, baryon density distributions obtained in UrQMD simulations with the PT EoS at $E_{\rm lab}=2A\,$GeV in the transverse $(x,y)$ coordinate plane
at $z = 0\,$fm.
The time was chosen as $t = 17$~fm/$c$  when  the central volume $V=27\,$fm$^3$ is inside the unstable region.
The projection of the volume $V$ on the $(x,y)$-plane is shown in Fig.~\ref{dens-coord} by the white squares. 
One observes that the volume $V$ with average baryon density $n_B=3.2\,n_0$ contains either a high~(liquid) or low~(gas) baryon density, or clump vs. no-clump scenario. These different states inside the central box change from one event to another, leading to large event-by-event fluctuations of the baryon density.
In the case of the CMF EoS no large changes of the baryon density inside the central box are observed. 

The UrQMD results for the baryon number distributions in the central box at $E_{lab}=2A\,$GeV are shown in 
Fig.~\ref{distr} (a) and (b), corresponding
to $n_B=3.2\,n_0$
and $n_B=1.2\,n_0$, respectively.
The probability to observe $B$ baryons $P(B)$ for the CMF equation of state can be described with a narrow, $\omega[B]<1$, bell shaped curve. In the case of the PT EoS the $P(B)$ distribution is much wider, $\omega[B]>1,$ and asymmetric around the mean. In order to emphasize the bimodal nature of this distribution we fit it by a sum of two Gaussian distributions. Note that $B\approx N_B$ and $N_{\overline{B}}\approx 0$ at $E_{\rm lab }=2A\,$GeV.
At the $n_B=1.2\,n_0$ a noticeable difference in the baryon number distributions is still seen
due to the memory effect of 
the dynamical expansion through the spinodal region for the PT EoS.
 
\subsection{Momentum space}

The experimental observations are limited to momenta of produced hadrons, therefore,
any information about spatial correlations has to be extracted from measurements performed in momentum space.
This is feasible in the presence of strong space-momentum correlations due to collective flow, in particular, due to Bjorken  longitudinal flow at high collision energies, see the recently developed
subensemble acceptance method
\cite{Vovchenko:2020tsr,Poberezhnyuk:2020ayn,Vovchenko:2020gne}. The method is not applicable in direct proximity to the critical point where correlation length reaches the size of the system and inside a mixed phase region where one can expect coexistence of two phases and strong non-equilibrium effects.
At lower collision energies, such as those considered in the present work, the collective flow is, however, 
rather weaker. Thus, new ways to extract signal of fluctuation enhancement are required.
Here, we compute the baryon number distribution in the momentum space by imposing rapidity cuts around midrapidity in the center of mass system of Au+Au.
To understand the effect of the existing coordinate space fluctuations on momentum space observations,
figures ~\ref{distr} (c) and (d) present the $p_t$-integrated baryon number distributions inside a $|y|<0.05$ rapidity interval. 
The times shown correspond to those already presented in Figs.~\ref{distr} (a) and (b) in the coordinate space. In contrast to the enhanced fluctuations in the coordinate space due to the PT, no qualitative difference between the $P(B)$ distributions with or without the PT is observed. In both the CMF and PT cases the distributions are single-peaked and close to the Poisson distribution. The suppression effects due to global charge conservation lead to $\omega[B]$ being slightly smaller than unity.

The fluctuation measures (\ref{omega}-\ref{kurt}) for the final protons calculated in the UrQMD model at $E_{\rm lab}=2\,A$GeV for $3~ million$ events are shown in  Fig.~\ref{protons-final} as functions of the rapidity acceptance interval $\Delta y$. The statistical errors for the moments are estimated using the Delta method~\cite{kendall1963advanced} using the open source package \texttt{sample-moments}~\cite{SMgithub}.

The scaled variance $\omega[{\rm p}]$ does not show large fluctuation effects. One finds $\omega[{\rm p}]\cong 1\div1.04$ at small $\Delta y\cong 0.2$  and it decreases monotonously at larger values of $\Delta y$. 
As a function of the rapidity acceptance interval $\omega[{\rm p}]$
it is only weakly influenced by the presence of the PT. In addition, for the CMF EoS  the values of $\omega[{\rm p}]$ are even slightly larger than those for the PT case. This surprising result can be understood as follows: In the collisions with the CMF-EoS a larger pressure gradient in the initial overlap zone is created which starts to create a more violent sidewards push of the participants into the spectator region. These participants then have a higher chance to rescatter with the remaining spectators. This increases the average number of participants in the CMF scenario as compared to the PT case. The enhancement of fluctuations in the CMF case is a result of the increased volume fluctuations which is a dynamical effect and is only indirectly related to the EoS. Furthermore, it can be observed that the cumulant ratios in both scenarios drop below the binomial baseline at large rapidities. This interesting observation is due to the rapidity dependence of iso-spin randomization (or pion production), where protons with large rapidity are more likely to stem from elastic scatterings.  Since elastic scatterings cannot change the isospin of the proton, the effect of isospin randomization is significantly suppressed at higher rapidities and thus the conservation effect dominates. 

The results for the skewness and
kurtosis shown in Figs.~\ref{protons-final} (b) and (c) are qualitatively very similar to the results for the scaled variance. The large negative values of the kurtosis again result from the large volume fluctuations with a strong contribution of conservation laws.
This indicates that the enhancement of fluctuations seen in the coordinate space is almost completely washed out when one turns to the momentum space. Momentum space fluctuations are dominated by the interplay between volume fluctuations and conservation laws which can act in a non-trivial way as a function of rapidity at these low beam energies.

\section{Conclusions}
\label{sec-5}

The baryon and proton number fluctuations in a presence of the first order phase transition (PT) were studied. 
We used the UrQMD-3.5 model with a density-dependent mean field interactions to simulate Au+Au collisions at intermediate energies.
Two different equations of state, the CMF and phase transition augmented CMF, were used.
By construction, the two EoS 
have the same properties at $n_B<2n_0$. 
However, they differ at higher baryon densities due to the presence of the PT. 
In both scenarios,  heavy-ion collisions follow very similar trajectories in the $(n_B,T)$ plane, the main difference being as expected a longer expansion time in the presence of the PT.

Despite similar trajectories on the phase diagram in the two scenarios, we observe notable differences in baryon number fluctuations in the central coordinate space volume.
The fluctuations exhibit an enhancement in the PT case, and it persists for some later stages of the collision when the system has already left the spinodal region.
The largest effect is observed at the collision energy of $E_{\rm lab}=2A\,$GeV which is purely related to the location of the unstable phase in the equation of state used.

On the other hand, 
the enhancement of proton number fluctuations due to the PT
is not seen when these fluctuations are analyzed in momentum space.
The scaled variance,  skewness, and kurtosis  of final state  protons as functions of the rapidity acceptance interval $\Delta y$ are not sensitive to the PT.

At large RHIC and LHC collision energies the A+A data demonstrate a monotonous decrease of proton number fluctuations  with $\Delta y$. This behavior is considered to be the consequence  of baryon conservation and excluded volume repulsion effects~\cite{Vovchenko:2022syc}. 
The qualitatively different behavior of increase of $\omega[{\rm p}]$ with $\Delta y$ from unity
up to large values of $\omega[{\rm p}]\cong 2.3$ at $\Delta y= 1$  was reported  by the HADES Collaboration in Au+Au collisions at $E_{\rm lab}=1.23A$~GeV~\cite{HADES:2020wpc}.

The present study shows that these large baryon number fluctuations are unlikely
to be 
the result of spinodal amplification due to a phase transition.
An alternative explanation of the HADES data
was suggested in the recent paper~\cite{Savchuk:2022ljy}. 
To search for interesting physics in the EoS, one should first exclude
the event-by-event fluctuations of nucleon participants. 
At small collision energy there
is a significant number of light nuclei and intermediate nuclear fragments in the final state.  
This can influence fluctuations of the number of bare protons.
Our UrQMD simulations do not yet include light nuclei and nuclear fragment production 
which could be relevant at intermediate collision energies \cite{https://doi.org/10.48550/arxiv.2208.11802,https://doi.org/10.48550/arxiv.2201.13374}.
We hope to address the question of the role of nuclear clusters for proton number fluctuations at HADES energy region in future studies.

	\begin{acknowledgments}
	The authors thank 
S. Pratt, M. Gazdzicki,
 V.~Kuznietsov, A. Rustamov, and H.~Stoecker for fruitful comments and discussions. O.S. acknowledges the scholarship grant from the GET$\_$INvolved Programme of FAIR/GSI and support by the Department of Energy Office of Science through grant no. DE-FG02-03ER41259. This work is supported by the National Academy of Sciences of Ukraine, Grant No.  0122U200259.  
 M.I.G. and R.V.P. acknowledge the support from the Alexander von Humboldt Foundation.
 A.M. acknowledges the Stern--Gerlach Postdoctoral fellowship of the Stiftung Polytechnische Gesellschaft.
 V.V. was supported through the U.S. Department of Energy, 
Office of Science, Office of Nuclear Physics, under contract number 
DE-FG02-00ER41132. 
	\end{acknowledgments}

\bibliography{main.bib}

\begin{thebibliography}{64}%
\makeatletter
\providecommand \@ifxundefined [1]{%
 \@ifx{#1\undefined}
}%
\providecommand \@ifnum [1]{%
 \ifnum #1\expandafter \@firstoftwo
 \else \expandafter \@secondoftwo
 \fi
}%
\providecommand \@ifx [1]{%
 \ifx #1\expandafter \@firstoftwo
 \else \expandafter \@secondoftwo
 \fi
}%
\providecommand \natexlab [1]{#1}%
\providecommand \enquote  [1]{``#1''}%
\providecommand \bibnamefont  [1]{#1}%
\providecommand \bibfnamefont [1]{#1}%
\providecommand \citenamefont [1]{#1}%
\providecommand \href@noop [0]{\@secondoftwo}%
\providecommand \href [0]{\begingroup \@sanitize@url \@href}%
\providecommand \@href[1]{\@@startlink{#1}\@@href}%
\providecommand \@@href[1]{\endgroup#1\@@endlink}%
\providecommand \@sanitize@url [0]{\catcode `\\12\catcode `\$12\catcode
  `\&12\catcode `\#12\catcode `\^12\catcode `\_12\catcode `\%12\relax}%
\providecommand \@@startlink[1]{}%
\providecommand \@@endlink[0]{}%
\providecommand \url  [0]{\begingroup\@sanitize@url \@url }%
\providecommand \@url [1]{\endgroup\@href {#1}{\urlprefix }}%
\providecommand \urlprefix  [0]{URL }%
\providecommand \Eprint [0]{\href }%
\providecommand \doibase [0]{http://dx.doi.org/}%
\providecommand \selectlanguage [0]{\@gobble}%
\providecommand \bibinfo  [0]{\@secondoftwo}%
\providecommand \bibfield  [0]{\@secondoftwo}%
\providecommand \translation [1]{[#1]}%
\providecommand \BibitemOpen [0]{}%
\providecommand \bibitemStop [0]{}%
\providecommand \bibitemNoStop [0]{.\EOS\space}%
\providecommand \EOS [0]{\spacefactor3000\relax}%
\providecommand \BibitemShut  [1]{\csname bibitem#1\endcsname}%
\let\auto@bib@innerbib\@empty
\bibitem [{\citenamefont {Busza}\ \emph {et~al.}(2018)\citenamefont {Busza},
  \citenamefont {Rajagopal},\ and\ \citenamefont {van~der
  Schee}}]{Busza:2018rrf}%
  \BibitemOpen
  \bibfield  {author} {\bibinfo {author} {\bibfnamefont {W.}~\bibnamefont
  {Busza}}, \bibinfo {author} {\bibfnamefont {K.}~\bibnamefont {Rajagopal}}, \
  and\ \bibinfo {author} {\bibfnamefont {W.}~\bibnamefont {van~der Schee}},\
  }\href {\doibase 10.1146/annurev-nucl-101917-020852} {\bibfield  {journal}
  {\bibinfo  {journal} {Ann. Rev. Nucl. Part. Sci.}\ }\textbf {\bibinfo
  {volume} {68}},\ \bibinfo {pages} {339} (\bibinfo {year} {2018})},\ \Eprint
  {http://arxiv.org/abs/1802.04801} {arXiv:1802.04801 [hep-ph]} \BibitemShut
  {NoStop}%
\bibitem [{\citenamefont {Bzdak}\ \emph {et~al.}(2020)\citenamefont {Bzdak},
  \citenamefont {Esumi}, \citenamefont {Koch}, \citenamefont {Liao},
  \citenamefont {Stephanov},\ and\ \citenamefont {Xu}}]{Bzdak:2019pkr}%
  \BibitemOpen
  \bibfield  {author} {\bibinfo {author} {\bibfnamefont {A.}~\bibnamefont
  {Bzdak}}, \bibinfo {author} {\bibfnamefont {S.}~\bibnamefont {Esumi}},
  \bibinfo {author} {\bibfnamefont {V.}~\bibnamefont {Koch}}, \bibinfo {author}
  {\bibfnamefont {J.}~\bibnamefont {Liao}}, \bibinfo {author} {\bibfnamefont
  {M.}~\bibnamefont {Stephanov}}, \ and\ \bibinfo {author} {\bibfnamefont
  {N.}~\bibnamefont {Xu}},\ }\href {\doibase 10.1016/j.physrep.2020.01.005}
  {\bibfield  {journal} {\bibinfo  {journal} {Phys. Rept.}\ }\textbf {\bibinfo
  {volume} {853}},\ \bibinfo {pages} {1} (\bibinfo {year} {2020})},\ \Eprint
  {http://arxiv.org/abs/1906.00936} {arXiv:1906.00936 [nucl-th]} \BibitemShut
  {NoStop}%
\bibitem [{\citenamefont {Lin}\ \emph {et~al.}(2005)\citenamefont {Lin},
  \citenamefont {Ko}, \citenamefont {Li}, \citenamefont {Zhang},\ and\
  \citenamefont {Pal}}]{PhysRevC.72.064901}%
  \BibitemOpen
  \bibfield  {author} {\bibinfo {author} {\bibfnamefont {Z.-W.}\ \bibnamefont
  {Lin}}, \bibinfo {author} {\bibfnamefont {C.~M.}\ \bibnamefont {Ko}},
  \bibinfo {author} {\bibfnamefont {B.-A.}\ \bibnamefont {Li}}, \bibinfo
  {author} {\bibfnamefont {B.}~\bibnamefont {Zhang}}, \ and\ \bibinfo {author}
  {\bibfnamefont {S.}~\bibnamefont {Pal}},\ }\href {\doibase
  10.1103/PhysRevC.72.064901} {\bibfield  {journal} {\bibinfo  {journal} {Phys.
  Rev. C}\ }\textbf {\bibinfo {volume} {72}},\ \bibinfo {pages} {064901}
  (\bibinfo {year} {2005})}\BibitemShut {NoStop}%
\bibitem [{\citenamefont {Petersen}\ \emph {et~al.}(2019)\citenamefont
  {Petersen}, \citenamefont {Oliinychenko}, \citenamefont {Mayer},
  \citenamefont {Staudenmaier},\ and\ \citenamefont {Ryu}}]{Petersen_2019}%
  \BibitemOpen
  \bibfield  {author} {\bibinfo {author} {\bibfnamefont {H.}~\bibnamefont
  {Petersen}}, \bibinfo {author} {\bibfnamefont {D.}~\bibnamefont
  {Oliinychenko}}, \bibinfo {author} {\bibfnamefont {M.}~\bibnamefont {Mayer}},
  \bibinfo {author} {\bibfnamefont {J.}~\bibnamefont {Staudenmaier}}, \ and\
  \bibinfo {author} {\bibfnamefont {S.}~\bibnamefont {Ryu}},\ }\href {\doibase
  10.1016/j.nuclphysa.2018.08.008} {\bibfield  {journal} {\bibinfo  {journal}
  {Nuclear Physics A}\ }\textbf {\bibinfo {volume} {982}},\ \bibinfo {pages}
  {399} (\bibinfo {year} {2019})}\BibitemShut {NoStop}%
\bibitem [{\citenamefont {Bass}\ \emph {et~al.}(1998)\citenamefont {Bass} \emph
  {et~al.}}]{Bass:1998ca}%
  \BibitemOpen
  \bibfield  {author} {\bibinfo {author} {\bibfnamefont {S.~A.}\ \bibnamefont
  {Bass}} \emph {et~al.},\ }\href {\doibase 10.1016/S0146-6410(98)00058-1}
  {\bibfield  {journal} {\bibinfo  {journal} {Prog. Part. Nucl. Phys.}\
  }\textbf {\bibinfo {volume} {41}},\ \bibinfo {pages} {255} (\bibinfo {year}
  {1998})},\ \Eprint {http://arxiv.org/abs/nucl-th/9803035}
  {arXiv:nucl-th/9803035} \BibitemShut {NoStop}%
\bibitem [{\citenamefont {Bleicher}\ \emph {et~al.}(1999)\citenamefont
  {Bleicher} \emph {et~al.}}]{Bleicher:1999xi}%
  \BibitemOpen
  \bibfield  {author} {\bibinfo {author} {\bibfnamefont {M.}~\bibnamefont
  {Bleicher}} \emph {et~al.},\ }\href {\doibase 10.1088/0954-3899/25/9/308}
  {\bibfield  {journal} {\bibinfo  {journal} {J. Phys. G}\ }\textbf {\bibinfo
  {volume} {25}},\ \bibinfo {pages} {1859} (\bibinfo {year} {1999})},\ \Eprint
  {http://arxiv.org/abs/hep-ph/9909407} {arXiv:hep-ph/9909407} \BibitemShut
  {NoStop}%
\bibitem [{\citenamefont {Bleicher}\ and\ \citenamefont
  {Bratkovskaya}(2022)}]{Bleicher:2022kcu}%
  \BibitemOpen
  \bibfield  {author} {\bibinfo {author} {\bibfnamefont {M.}~\bibnamefont
  {Bleicher}}\ and\ \bibinfo {author} {\bibfnamefont {E.}~\bibnamefont
  {Bratkovskaya}},\ }\href {\doibase 10.1016/j.ppnp.2021.103920} {\bibfield
  {journal} {\bibinfo  {journal} {Prog. Part. Nucl. Phys.}\ }\textbf {\bibinfo
  {volume} {122}},\ \bibinfo {pages} {103920} (\bibinfo {year}
  {2022})}\BibitemShut {NoStop}%
\bibitem [{\citenamefont {Li}\ and\ \citenamefont {Ko}(2017)}]{Li:2016uvu}%
  \BibitemOpen
  \bibfield  {author} {\bibinfo {author} {\bibfnamefont {F.}~\bibnamefont
  {Li}}\ and\ \bibinfo {author} {\bibfnamefont {C.~M.}\ \bibnamefont {Ko}},\
  }\href {\doibase 10.1103/PhysRevC.95.055203} {\bibfield  {journal} {\bibinfo
  {journal} {Phys. Rev. C}\ }\textbf {\bibinfo {volume} {95}},\ \bibinfo
  {pages} {055203} (\bibinfo {year} {2017})},\ \Eprint
  {http://arxiv.org/abs/1606.05012} {arXiv:1606.05012 [nucl-th]} \BibitemShut
  {NoStop}%
\bibitem [{\citenamefont {Steinheimer}\ \emph {et~al.}(2018)\citenamefont
  {Steinheimer}, \citenamefont {Wang}, \citenamefont {Mukherjee}, \citenamefont
  {Ye}, \citenamefont {Guo}, \citenamefont {Li},\ and\ \citenamefont
  {Stoecker}}]{Steinheimer:2018rnd}%
  \BibitemOpen
  \bibfield  {author} {\bibinfo {author} {\bibfnamefont {J.}~\bibnamefont
  {Steinheimer}}, \bibinfo {author} {\bibfnamefont {Y.}~\bibnamefont {Wang}},
  \bibinfo {author} {\bibfnamefont {A.}~\bibnamefont {Mukherjee}}, \bibinfo
  {author} {\bibfnamefont {Y.}~\bibnamefont {Ye}}, \bibinfo {author}
  {\bibfnamefont {C.}~\bibnamefont {Guo}}, \bibinfo {author} {\bibfnamefont
  {Q.}~\bibnamefont {Li}}, \ and\ \bibinfo {author} {\bibfnamefont
  {H.}~\bibnamefont {Stoecker}},\ }\href {\doibase
  10.1016/j.physletb.2018.07.068} {\bibfield  {journal} {\bibinfo  {journal}
  {Phys. Lett. B}\ }\textbf {\bibinfo {volume} {785}},\ \bibinfo {pages} {40}
  (\bibinfo {year} {2018})},\ \Eprint {http://arxiv.org/abs/1804.08936}
  {arXiv:1804.08936 [nucl-th]} \BibitemShut {NoStop}%
\bibitem [{\citenamefont {Ohnishi}\ and\ \citenamefont
  {Randrup}(1995)}]{PhysRevLett.75.596}%
  \BibitemOpen
  \bibfield  {author} {\bibinfo {author} {\bibfnamefont {A.}~\bibnamefont
  {Ohnishi}}\ and\ \bibinfo {author} {\bibfnamefont {J.}~\bibnamefont
  {Randrup}},\ }\href {\doibase 10.1103/PhysRevLett.75.596} {\bibfield
  {journal} {\bibinfo  {journal} {Phys. Rev. Lett.}\ }\textbf {\bibinfo
  {volume} {75}},\ \bibinfo {pages} {596} (\bibinfo {year} {1995})}\BibitemShut
  {NoStop}%
\bibitem [{\citenamefont {Rischke}\ \emph {et~al.}(1995)\citenamefont
  {Rischke}, \citenamefont {P\"urs\"un}, \citenamefont {Maruhn}, \citenamefont
  {Stoecker},\ and\ \citenamefont {Greiner}}]{Rischke:1995pe}%
  \BibitemOpen
  \bibfield  {author} {\bibinfo {author} {\bibfnamefont {D.~H.}\ \bibnamefont
  {Rischke}}, \bibinfo {author} {\bibfnamefont {Y.}~\bibnamefont {P\"urs\"un}},
  \bibinfo {author} {\bibfnamefont {J.~A.}\ \bibnamefont {Maruhn}}, \bibinfo
  {author} {\bibfnamefont {H.}~\bibnamefont {Stoecker}}, \ and\ \bibinfo
  {author} {\bibfnamefont {W.}~\bibnamefont {Greiner}},\ }\href {\doibase
  10.1007/BF03053749} {\bibfield  {journal} {\bibinfo  {journal} {Acta Phys.
  Hung. A}\ }\textbf {\bibinfo {volume} {1}},\ \bibinfo {pages} {309} (\bibinfo
  {year} {1995})},\ \Eprint {http://arxiv.org/abs/nucl-th/9505014}
  {arXiv:nucl-th/9505014} \BibitemShut {NoStop}%
\bibitem [{\citenamefont {Stoecker}(2005)}]{Stoecker:2004qu}%
  \BibitemOpen
  \bibfield  {author} {\bibinfo {author} {\bibfnamefont {H.}~\bibnamefont
  {Stoecker}},\ }\href {\doibase 10.1016/j.nuclphysa.2004.12.074} {\bibfield
  {journal} {\bibinfo  {journal} {Nucl. Phys. A}\ }\textbf {\bibinfo {volume}
  {750}},\ \bibinfo {pages} {121} (\bibinfo {year} {2005})},\ \Eprint
  {http://arxiv.org/abs/nucl-th/0406018} {arXiv:nucl-th/0406018} \BibitemShut
  {NoStop}%
\bibitem [{\citenamefont {Brachmann}\ \emph
  {et~al.}(2000{\natexlab{a}})\citenamefont {Brachmann}, \citenamefont {Soff},
  \citenamefont {Dumitru}, \citenamefont {Stoecker}, \citenamefont {Maruhn},
  \citenamefont {Greiner}, \citenamefont {Bravina},\ and\ \citenamefont
  {Rischke}}]{Brachmann:1999xt}%
  \BibitemOpen
  \bibfield  {author} {\bibinfo {author} {\bibfnamefont {J.}~\bibnamefont
  {Brachmann}}, \bibinfo {author} {\bibfnamefont {S.}~\bibnamefont {Soff}},
  \bibinfo {author} {\bibfnamefont {A.}~\bibnamefont {Dumitru}}, \bibinfo
  {author} {\bibfnamefont {H.}~\bibnamefont {Stoecker}}, \bibinfo {author}
  {\bibfnamefont {J.~A.}\ \bibnamefont {Maruhn}}, \bibinfo {author}
  {\bibfnamefont {W.}~\bibnamefont {Greiner}}, \bibinfo {author} {\bibfnamefont
  {L.~V.}\ \bibnamefont {Bravina}}, \ and\ \bibinfo {author} {\bibfnamefont
  {D.~H.}\ \bibnamefont {Rischke}},\ }\href {\doibase
  10.1103/PhysRevC.61.024909} {\bibfield  {journal} {\bibinfo  {journal} {Phys.
  Rev. C}\ }\textbf {\bibinfo {volume} {61}},\ \bibinfo {pages} {024909}
  (\bibinfo {year} {2000}{\natexlab{a}})},\ \Eprint
  {http://arxiv.org/abs/nucl-th/9908010} {arXiv:nucl-th/9908010} \BibitemShut
  {NoStop}%
\bibitem [{\citenamefont {Brachmann}\ \emph
  {et~al.}(2000{\natexlab{b}})\citenamefont {Brachmann}, \citenamefont
  {Dumitru}, \citenamefont {Stoecker},\ and\ \citenamefont
  {Greiner}}]{Brachmann:1999mp}%
  \BibitemOpen
  \bibfield  {author} {\bibinfo {author} {\bibfnamefont {J.}~\bibnamefont
  {Brachmann}}, \bibinfo {author} {\bibfnamefont {A.}~\bibnamefont {Dumitru}},
  \bibinfo {author} {\bibfnamefont {H.}~\bibnamefont {Stoecker}}, \ and\
  \bibinfo {author} {\bibfnamefont {W.}~\bibnamefont {Greiner}},\ }\href
  {\doibase 10.1007/s100500070077} {\bibfield  {journal} {\bibinfo  {journal}
  {Eur. Phys. J. A}\ }\textbf {\bibinfo {volume} {8}},\ \bibinfo {pages} {549}
  (\bibinfo {year} {2000}{\natexlab{b}})},\ \Eprint
  {http://arxiv.org/abs/nucl-th/9912014} {arXiv:nucl-th/9912014} \BibitemShut
  {NoStop}%
\bibitem [{\citenamefont {Oliinychenko}\ \emph {et~al.}(2022)\citenamefont
  {Oliinychenko}, \citenamefont {Sorensen}, \citenamefont {Koch},\ and\
  \citenamefont {McLerran}}]{Oliinychenko:2022uvy}%
  \BibitemOpen
  \bibfield  {author} {\bibinfo {author} {\bibfnamefont {D.}~\bibnamefont
  {Oliinychenko}}, \bibinfo {author} {\bibfnamefont {A.}~\bibnamefont
  {Sorensen}}, \bibinfo {author} {\bibfnamefont {V.}~\bibnamefont {Koch}}, \
  and\ \bibinfo {author} {\bibfnamefont {L.}~\bibnamefont {McLerran}},\
  }\href@noop {} {\  (\bibinfo {year} {2022})},\ \Eprint
  {http://arxiv.org/abs/2208.11996} {arXiv:2208.11996 [nucl-th]} \BibitemShut
  {NoStop}%
\bibitem [{\citenamefont {Stephanov}\ \emph {et~al.}(1998)\citenamefont
  {Stephanov}, \citenamefont {Rajagopal},\ and\ \citenamefont
  {Shuryak}}]{Stephanov:1998dy}%
  \BibitemOpen
  \bibfield  {author} {\bibinfo {author} {\bibfnamefont {M.~A.}\ \bibnamefont
  {Stephanov}}, \bibinfo {author} {\bibfnamefont {K.}~\bibnamefont
  {Rajagopal}}, \ and\ \bibinfo {author} {\bibfnamefont {E.~V.}\ \bibnamefont
  {Shuryak}},\ }\href {\doibase 10.1103/PhysRevLett.81.4816} {\bibfield
  {journal} {\bibinfo  {journal} {Phys. Rev. Lett.}\ }\textbf {\bibinfo
  {volume} {81}},\ \bibinfo {pages} {4816} (\bibinfo {year} {1998})},\ \Eprint
  {http://arxiv.org/abs/hep-ph/9806219} {arXiv:hep-ph/9806219 [hep-ph]}
  \BibitemShut {NoStop}%
\bibitem [{\citenamefont {Stephanov}\ \emph {et~al.}(1999)\citenamefont
  {Stephanov}, \citenamefont {Rajagopal},\ and\ \citenamefont
  {Shuryak}}]{Stephanov:1999zu}%
  \BibitemOpen
  \bibfield  {author} {\bibinfo {author} {\bibfnamefont {M.~A.}\ \bibnamefont
  {Stephanov}}, \bibinfo {author} {\bibfnamefont {K.}~\bibnamefont
  {Rajagopal}}, \ and\ \bibinfo {author} {\bibfnamefont {E.~V.}\ \bibnamefont
  {Shuryak}},\ }\href {\doibase 10.1103/PhysRevD.60.114028} {\bibfield
  {journal} {\bibinfo  {journal} {Phys. Rev.}\ }\textbf {\bibinfo {volume}
  {D60}},\ \bibinfo {pages} {114028} (\bibinfo {year} {1999})},\ \Eprint
  {http://arxiv.org/abs/hep-ph/9903292} {arXiv:hep-ph/9903292 [hep-ph]}
  \BibitemShut {NoStop}%
\bibitem [{\citenamefont {Jeon}\ and\ \citenamefont
  {Koch}(2000)}]{Jeon:2000wg}%
  \BibitemOpen
  \bibfield  {author} {\bibinfo {author} {\bibfnamefont {S.}~\bibnamefont
  {Jeon}}\ and\ \bibinfo {author} {\bibfnamefont {V.}~\bibnamefont {Koch}},\
  }\href {\doibase 10.1103/PhysRevLett.85.2076} {\bibfield  {journal} {\bibinfo
   {journal} {Phys. Rev. Lett.}\ }\textbf {\bibinfo {volume} {85}},\ \bibinfo
  {pages} {2076} (\bibinfo {year} {2000})},\ \Eprint
  {http://arxiv.org/abs/hep-ph/0003168} {arXiv:hep-ph/0003168 [hep-ph]}
  \BibitemShut {NoStop}%
\bibitem [{\citenamefont {Asakawa}\ \emph {et~al.}(2000)\citenamefont
  {Asakawa}, \citenamefont {Heinz},\ and\ \citenamefont
  {Muller}}]{Asakawa:2000wh}%
  \BibitemOpen
  \bibfield  {author} {\bibinfo {author} {\bibfnamefont {M.}~\bibnamefont
  {Asakawa}}, \bibinfo {author} {\bibfnamefont {U.~W.}\ \bibnamefont {Heinz}},
  \ and\ \bibinfo {author} {\bibfnamefont {B.}~\bibnamefont {Muller}},\ }\href
  {\doibase 10.1103/PhysRevLett.85.2072} {\bibfield  {journal} {\bibinfo
  {journal} {Phys. Rev. Lett.}\ }\textbf {\bibinfo {volume} {85}},\ \bibinfo
  {pages} {2072} (\bibinfo {year} {2000})},\ \Eprint
  {http://arxiv.org/abs/hep-ph/0003169} {arXiv:hep-ph/0003169 [hep-ph]}
  \BibitemShut {NoStop}%
\bibitem [{\citenamefont {Sun}\ \emph {et~al.}(2017)\citenamefont {Sun},
  \citenamefont {Chen}, \citenamefont {Ko},\ and\ \citenamefont
  {Xu}}]{Sun:2017xrx}%
  \BibitemOpen
  \bibfield  {author} {\bibinfo {author} {\bibfnamefont {K.-J.}\ \bibnamefont
  {Sun}}, \bibinfo {author} {\bibfnamefont {L.-W.}\ \bibnamefont {Chen}},
  \bibinfo {author} {\bibfnamefont {C.~M.}\ \bibnamefont {Ko}}, \ and\ \bibinfo
  {author} {\bibfnamefont {Z.}~\bibnamefont {Xu}},\ }\href {\doibase
  10.1016/j.physletb.2017.09.056} {\bibfield  {journal} {\bibinfo  {journal}
  {Phys. Lett. B}\ }\textbf {\bibinfo {volume} {774}},\ \bibinfo {pages} {103}
  (\bibinfo {year} {2017})},\ \Eprint {http://arxiv.org/abs/1702.07620}
  {arXiv:1702.07620 [nucl-th]} \BibitemShut {NoStop}%
\bibitem [{\citenamefont {Shuryak}(1978)}]{Shuryak:1978ij}%
  \BibitemOpen
  \bibfield  {author} {\bibinfo {author} {\bibfnamefont {E.~V.}\ \bibnamefont
  {Shuryak}},\ }\href {\doibase 10.1016/0370-2693(78)90370-2} {\bibfield
  {journal} {\bibinfo  {journal} {Phys. Lett. B}\ }\textbf {\bibinfo {volume}
  {78}},\ \bibinfo {pages} {150} (\bibinfo {year} {1978})}\BibitemShut
  {NoStop}%
\bibitem [{\citenamefont {McLerran}\ and\ \citenamefont
  {Toimela}(1985)}]{McLerran:1984ay}%
  \BibitemOpen
  \bibfield  {author} {\bibinfo {author} {\bibfnamefont {L.~D.}\ \bibnamefont
  {McLerran}}\ and\ \bibinfo {author} {\bibfnamefont {T.}~\bibnamefont
  {Toimela}},\ }\href {\doibase 10.1103/PhysRevD.31.545} {\bibfield  {journal}
  {\bibinfo  {journal} {Phys. Rev. D}\ }\textbf {\bibinfo {volume} {31}},\
  \bibinfo {pages} {545} (\bibinfo {year} {1985})}\BibitemShut {NoStop}%
\bibitem [{\citenamefont {Bratkovskaya}\ and\ \citenamefont
  {Cassing}(1997)}]{Bratkovskaya:1996qe}%
  \BibitemOpen
  \bibfield  {author} {\bibinfo {author} {\bibfnamefont {E.~L.}\ \bibnamefont
  {Bratkovskaya}}\ and\ \bibinfo {author} {\bibfnamefont {W.}~\bibnamefont
  {Cassing}},\ }\href {\doibase 10.1016/S0375-9474(97)00140-1} {\bibfield
  {journal} {\bibinfo  {journal} {Nucl. Phys. A}\ }\textbf {\bibinfo {volume}
  {619}},\ \bibinfo {pages} {413} (\bibinfo {year} {1997})},\ \Eprint
  {http://arxiv.org/abs/nucl-th/9611042} {arXiv:nucl-th/9611042} \BibitemShut
  {NoStop}%
\bibitem [{\citenamefont {Rapp}\ and\ \citenamefont
  {Wambach}(2000)}]{Rapp:1999ej}%
  \BibitemOpen
  \bibfield  {author} {\bibinfo {author} {\bibfnamefont {R.}~\bibnamefont
  {Rapp}}\ and\ \bibinfo {author} {\bibfnamefont {J.}~\bibnamefont {Wambach}},\
  }\href {\doibase 10.1007/0-306-47101-9_1} {\bibfield  {journal} {\bibinfo
  {journal} {Adv. Nucl. Phys.}\ }\textbf {\bibinfo {volume} {25}},\ \bibinfo
  {pages} {1} (\bibinfo {year} {2000})},\ \Eprint
  {http://arxiv.org/abs/hep-ph/9909229} {arXiv:hep-ph/9909229} \BibitemShut
  {NoStop}%
\bibitem [{\citenamefont {Savchuk}\ \emph
  {et~al.}(2022{\natexlab{a}})\citenamefont {Savchuk}, \citenamefont
  {Motornenko}, \citenamefont {Steinheimer}, \citenamefont {Vovchenko},
  \citenamefont {Bleicher}, \citenamefont {Gorenstein},\ and\ \citenamefont
  {Galatyuk}}]{Savchuk:2022aev}%
  \BibitemOpen
  \bibfield  {author} {\bibinfo {author} {\bibfnamefont {O.}~\bibnamefont
  {Savchuk}}, \bibinfo {author} {\bibfnamefont {A.}~\bibnamefont {Motornenko}},
  \bibinfo {author} {\bibfnamefont {J.}~\bibnamefont {Steinheimer}}, \bibinfo
  {author} {\bibfnamefont {V.}~\bibnamefont {Vovchenko}}, \bibinfo {author}
  {\bibfnamefont {M.}~\bibnamefont {Bleicher}}, \bibinfo {author}
  {\bibfnamefont {M.}~\bibnamefont {Gorenstein}}, \ and\ \bibinfo {author}
  {\bibfnamefont {T.}~\bibnamefont {Galatyuk}},\ }\href@noop {} {\  (\bibinfo
  {year} {2022}{\natexlab{a}})},\ \Eprint {http://arxiv.org/abs/2209.05267}
  {arXiv:2209.05267 [nucl-th]} \BibitemShut {NoStop}%
\bibitem [{\citenamefont {Adamczyk}\ \emph {et~al.}(2017)\citenamefont
  {Adamczyk} \emph {et~al.}}]{STAR:2017sal}%
  \BibitemOpen
  \bibfield  {author} {\bibinfo {author} {\bibfnamefont {L.}~\bibnamefont
  {Adamczyk}} \emph {et~al.} (\bibinfo {collaboration} {STAR}),\ }\href
  {\doibase 10.1103/PhysRevC.96.044904} {\bibfield  {journal} {\bibinfo
  {journal} {Phys. Rev. C}\ }\textbf {\bibinfo {volume} {96}},\ \bibinfo
  {pages} {044904} (\bibinfo {year} {2017})},\ \Eprint
  {http://arxiv.org/abs/1701.07065} {arXiv:1701.07065 [nucl-ex]} \BibitemShut
  {NoStop}%
\bibitem [{\citenamefont {Harabasz}\ \emph {et~al.}(2020)\citenamefont
  {Harabasz}, \citenamefont {Florkowski}, \citenamefont {Galatyuk},
  \citenamefont {Ma~Lgorzata~Gumberidze}, \citenamefont {Ryblewski},
  \citenamefont {Salabura},\ and\ \citenamefont {Stroth}}]{Harabasz:2020sei}%
  \BibitemOpen
  \bibfield  {author} {\bibinfo {author} {\bibfnamefont {S.}~\bibnamefont
  {Harabasz}}, \bibinfo {author} {\bibfnamefont {W.}~\bibnamefont
  {Florkowski}}, \bibinfo {author} {\bibfnamefont {T.}~\bibnamefont
  {Galatyuk}}, \bibinfo {author} {\bibfnamefont {t.}~\bibnamefont
  {Ma~Lgorzata~Gumberidze}}, \bibinfo {author} {\bibfnamefont {R.}~\bibnamefont
  {Ryblewski}}, \bibinfo {author} {\bibfnamefont {P.}~\bibnamefont {Salabura}},
  \ and\ \bibinfo {author} {\bibfnamefont {J.}~\bibnamefont {Stroth}},\ }\href
  {\doibase 10.1103/PhysRevC.102.054903} {\bibfield  {journal} {\bibinfo
  {journal} {Phys. Rev. C}\ }\textbf {\bibinfo {volume} {102}},\ \bibinfo
  {pages} {054903} (\bibinfo {year} {2020})},\ \Eprint
  {http://arxiv.org/abs/2003.12992} {arXiv:2003.12992 [nucl-th]} \BibitemShut
  {NoStop}%
\bibitem [{\citenamefont {Motornenko}\ \emph {et~al.}(2021)\citenamefont
  {Motornenko}, \citenamefont {Steinheimer}, \citenamefont {Vovchenko},
  \citenamefont {Stock},\ and\ \citenamefont {Stoecker}}]{Motornenko:2021nds}%
  \BibitemOpen
  \bibfield  {author} {\bibinfo {author} {\bibfnamefont {A.}~\bibnamefont
  {Motornenko}}, \bibinfo {author} {\bibfnamefont {J.}~\bibnamefont
  {Steinheimer}}, \bibinfo {author} {\bibfnamefont {V.}~\bibnamefont
  {Vovchenko}}, \bibinfo {author} {\bibfnamefont {R.}~\bibnamefont {Stock}}, \
  and\ \bibinfo {author} {\bibfnamefont {H.}~\bibnamefont {Stoecker}},\ }\href
  {\doibase 10.1016/j.physletb.2021.136703} {\bibfield  {journal} {\bibinfo
  {journal} {Phys. Lett. B}\ }\textbf {\bibinfo {volume} {822}},\ \bibinfo
  {pages} {136703} (\bibinfo {year} {2021})},\ \Eprint
  {http://arxiv.org/abs/2104.06036} {arXiv:2104.06036 [hep-ph]} \BibitemShut
  {NoStop}%
\bibitem [{\citenamefont {Bumnedpan}\ \emph {et~al.}(2022)\citenamefont
  {Bumnedpan}, \citenamefont {Steinheimer}, \citenamefont {Bleicher},
  \citenamefont {Limphirat},\ and\ \citenamefont {Herold}}]{Bumnedpan:2022lma}%
  \BibitemOpen
  \bibfield  {author} {\bibinfo {author} {\bibfnamefont {T.}~\bibnamefont
  {Bumnedpan}}, \bibinfo {author} {\bibfnamefont {J.}~\bibnamefont
  {Steinheimer}}, \bibinfo {author} {\bibfnamefont {M.}~\bibnamefont
  {Bleicher}}, \bibinfo {author} {\bibfnamefont {A.}~\bibnamefont {Limphirat}},
  \ and\ \bibinfo {author} {\bibfnamefont {C.}~\bibnamefont {Herold}},\ }\href
  {\doibase 10.1016/j.physletb.2022.137537} {\bibfield  {journal} {\bibinfo
  {journal} {Phys. Lett. B}\ }\textbf {\bibinfo {volume} {835}},\ \bibinfo
  {pages} {137537} (\bibinfo {year} {2022})},\ \Eprint
  {http://arxiv.org/abs/2209.04096} {arXiv:2209.04096 [nucl-th]} \BibitemShut
  {NoStop}%
\bibitem [{\citenamefont {Luo}\ and\ \citenamefont {Xu}(2017)}]{Luo:2017faz}%
  \BibitemOpen
  \bibfield  {author} {\bibinfo {author} {\bibfnamefont {X.}~\bibnamefont
  {Luo}}\ and\ \bibinfo {author} {\bibfnamefont {N.}~\bibnamefont {Xu}},\
  }\href {\doibase 10.1007/s41365-017-0257-0} {\bibfield  {journal} {\bibinfo
  {journal} {Nucl. Sci. Tech.}\ }\textbf {\bibinfo {volume} {28}},\ \bibinfo
  {pages} {112} (\bibinfo {year} {2017})},\ \Eprint
  {http://arxiv.org/abs/1701.02105} {arXiv:1701.02105 [nucl-ex]} \BibitemShut
  {NoStop}%
\bibitem [{\citenamefont {Steinheimer}\ and\ \citenamefont
  {Randrup}(2012)}]{Steinheimer:2012gc}%
  \BibitemOpen
  \bibfield  {author} {\bibinfo {author} {\bibfnamefont {J.}~\bibnamefont
  {Steinheimer}}\ and\ \bibinfo {author} {\bibfnamefont {J.}~\bibnamefont
  {Randrup}},\ }\href {\doibase 10.1103/PhysRevLett.109.212301} {\bibfield
  {journal} {\bibinfo  {journal} {Phys. Rev. Lett.}\ }\textbf {\bibinfo
  {volume} {109}},\ \bibinfo {pages} {212301} (\bibinfo {year} {2012})},\
  \Eprint {http://arxiv.org/abs/1209.2462} {arXiv:1209.2462 [nucl-th]}
  \BibitemShut {NoStop}%
\bibitem [{\citenamefont {Steinheimer}\ and\ \citenamefont
  {Randrup}(2013)}]{Steinheimer:2013glaa}%
  \BibitemOpen
  \bibfield  {author} {\bibinfo {author} {\bibfnamefont {J.}~\bibnamefont
  {Steinheimer}}\ and\ \bibinfo {author} {\bibfnamefont {J.}~\bibnamefont
  {Randrup}},\ }\href {\doibase 10.1103/PhysRevC.87.054903} {\bibfield
  {journal} {\bibinfo  {journal} {Phys. Rev. C}\ }\textbf {\bibinfo {volume}
  {87}},\ \bibinfo {pages} {054903} (\bibinfo {year} {2013})},\ \Eprint
  {http://arxiv.org/abs/1302.2956} {arXiv:1302.2956 [nucl-th]} \BibitemShut
  {NoStop}%
\bibitem [{\citenamefont {Berdnikov}\ and\ \citenamefont
  {Rajagopal}(2000)}]{Berdnikov_2000}%
  \BibitemOpen
  \bibfield  {author} {\bibinfo {author} {\bibfnamefont {B.}~\bibnamefont
  {Berdnikov}}\ and\ \bibinfo {author} {\bibfnamefont {K.}~\bibnamefont
  {Rajagopal}},\ }\href {\doibase 10.1103/physrevd.61.105017} {\bibfield
  {journal} {\bibinfo  {journal} {Physical Review D}\ }\textbf {\bibinfo
  {volume} {61}} (\bibinfo {year} {2000}),\
  10.1103/physrevd.61.105017}\BibitemShut {NoStop}%
\bibitem [{\citenamefont {Bluhm}\ \emph {et~al.}(2020)\citenamefont {Bluhm}
  \emph {et~al.}}]{BLUHM2020122016}%
  \BibitemOpen
  \bibfield  {author} {\bibinfo {author} {\bibfnamefont {M.}~\bibnamefont
  {Bluhm}} \emph {et~al.},\ }\href {\doibase 10.1016/j.nuclphysa.2020.122016}
  {\bibfield  {journal} {\bibinfo  {journal} {Nucl. Phys. A}\ }\textbf
  {\bibinfo {volume} {1003}},\ \bibinfo {pages} {122016} (\bibinfo {year}
  {2020})},\ \Eprint {http://arxiv.org/abs/2001.08831} {arXiv:2001.08831
  [nucl-th]} \BibitemShut {NoStop}%
\bibitem [{\citenamefont {Kitazawa}\ and\ \citenamefont
  {Asakawa}(2012)}]{Kitazawa:2012at}%
  \BibitemOpen
  \bibfield  {author} {\bibinfo {author} {\bibfnamefont {M.}~\bibnamefont
  {Kitazawa}}\ and\ \bibinfo {author} {\bibfnamefont {M.}~\bibnamefont
  {Asakawa}},\ }\href {\doibase 10.1103/PhysRevC.86.024904} {\bibfield
  {journal} {\bibinfo  {journal} {Phys. Rev. C}\ }\textbf {\bibinfo {volume}
  {86}},\ \bibinfo {pages} {024904} (\bibinfo {year} {2012})},\ \bibinfo {note}
  {[Erratum: Phys.Rev.C 86, 069902 (2012)]},\ \Eprint
  {http://arxiv.org/abs/1205.3292} {arXiv:1205.3292 [nucl-th]} \BibitemShut
  {NoStop}%
\bibitem [{\citenamefont {Mukherjee}\ \emph {et~al.}(2015)\citenamefont
  {Mukherjee}, \citenamefont {Venugopalan},\ and\ \citenamefont
  {Yin}}]{Mukherjee:2015swa}%
  \BibitemOpen
  \bibfield  {author} {\bibinfo {author} {\bibfnamefont {S.}~\bibnamefont
  {Mukherjee}}, \bibinfo {author} {\bibfnamefont {R.}~\bibnamefont
  {Venugopalan}}, \ and\ \bibinfo {author} {\bibfnamefont {Y.}~\bibnamefont
  {Yin}},\ }\href {\doibase 10.1103/PhysRevC.92.034912} {\bibfield  {journal}
  {\bibinfo  {journal} {Phys. Rev. C}\ }\textbf {\bibinfo {volume} {92}},\
  \bibinfo {pages} {034912} (\bibinfo {year} {2015})},\ \Eprint
  {http://arxiv.org/abs/1506.00645} {arXiv:1506.00645 [hep-ph]} \BibitemShut
  {NoStop}%
\bibitem [{\citenamefont {Pradeep}\ \emph {et~al.}(2022)\citenamefont
  {Pradeep}, \citenamefont {Rajagopal}, \citenamefont {Stephanov},\ and\
  \citenamefont {Yin}}]{Pradeep:2022mkf}%
  \BibitemOpen
  \bibfield  {author} {\bibinfo {author} {\bibfnamefont {M.}~\bibnamefont
  {Pradeep}}, \bibinfo {author} {\bibfnamefont {K.}~\bibnamefont {Rajagopal}},
  \bibinfo {author} {\bibfnamefont {M.}~\bibnamefont {Stephanov}}, \ and\
  \bibinfo {author} {\bibfnamefont {Y.}~\bibnamefont {Yin}},\ }\href {\doibase
  10.1103/PhysRevD.106.036017} {\bibfield  {journal} {\bibinfo  {journal}
  {Phys. Rev. D}\ }\textbf {\bibinfo {volume} {106}},\ \bibinfo {pages}
  {036017} (\bibinfo {year} {2022})},\ \Eprint
  {http://arxiv.org/abs/2204.00639} {arXiv:2204.00639 [hep-ph]} \BibitemShut
  {NoStop}%
\bibitem [{\citenamefont {Bravina}\ \emph {et~al.}(1998)\citenamefont {Bravina}
  \emph {et~al.}}]{Bravina:1998pi}%
  \BibitemOpen
  \bibfield  {author} {\bibinfo {author} {\bibfnamefont {L.~V.}\ \bibnamefont
  {Bravina}} \emph {et~al.},\ }\href {\doibase 10.1016/S0370-2693(98)00624-8}
  {\bibfield  {journal} {\bibinfo  {journal} {Phys. Lett. B}\ }\textbf
  {\bibinfo {volume} {434}},\ \bibinfo {pages} {379} (\bibinfo {year}
  {1998})},\ \Eprint {http://arxiv.org/abs/nucl-th/9804008}
  {arXiv:nucl-th/9804008} \BibitemShut {NoStop}%
\bibitem [{\citenamefont {Zabrodin}\ \emph {et~al.}(2009)\citenamefont
  {Zabrodin} \emph {et~al.}}]{Zabrodin:2009fz}%
  \BibitemOpen
  \bibfield  {author} {\bibinfo {author} {\bibfnamefont {E.~E.}\ \bibnamefont
  {Zabrodin}} \emph {et~al.},\ }\href {\doibase 10.1088/0954-3899/36/6/064065}
  {\bibfield  {journal} {\bibinfo  {journal} {J. Phys. G}\ }\textbf {\bibinfo
  {volume} {36}},\ \bibinfo {pages} {064065} (\bibinfo {year} {2009})},\
  \Eprint {http://arxiv.org/abs/0902.4601} {arXiv:0902.4601 [hep-ph]}
  \BibitemShut {NoStop}%
\bibitem [{\citenamefont {Omana~Kuttan}\ \emph {et~al.}(2022)\citenamefont
  {Omana~Kuttan}, \citenamefont {Motornenko}, \citenamefont {Steinheimer},
  \citenamefont {Stoecker}, \citenamefont {Nara},\ and\ \citenamefont
  {Bleicher}}]{OmanaKuttan:2022the}%
  \BibitemOpen
  \bibfield  {author} {\bibinfo {author} {\bibfnamefont {M.}~\bibnamefont
  {Omana~Kuttan}}, \bibinfo {author} {\bibfnamefont {A.}~\bibnamefont
  {Motornenko}}, \bibinfo {author} {\bibfnamefont {J.}~\bibnamefont
  {Steinheimer}}, \bibinfo {author} {\bibfnamefont {H.}~\bibnamefont
  {Stoecker}}, \bibinfo {author} {\bibfnamefont {Y.}~\bibnamefont {Nara}}, \
  and\ \bibinfo {author} {\bibfnamefont {M.}~\bibnamefont {Bleicher}},\ }\href
  {\doibase 10.1140/epjc/s10052-022-10400-2} {\bibfield  {journal} {\bibinfo
  {journal} {Eur. Phys. J. C}\ }\textbf {\bibinfo {volume} {82}},\ \bibinfo
  {pages} {427} (\bibinfo {year} {2022})},\ \Eprint
  {http://arxiv.org/abs/2201.01622} {arXiv:2201.01622 [nucl-th]} \BibitemShut
  {NoStop}%
\bibitem [{\citenamefont {Steinheimer}\ \emph {et~al.}(2022)\citenamefont
  {Steinheimer}, \citenamefont {Motornenko}, \citenamefont {Sorensen},
  \citenamefont {Nara}, \citenamefont {Koch},\ and\ \citenamefont
  {Bleicher}}]{Steinheimer:2022gqb}%
  \BibitemOpen
  \bibfield  {author} {\bibinfo {author} {\bibfnamefont {J.}~\bibnamefont
  {Steinheimer}}, \bibinfo {author} {\bibfnamefont {A.}~\bibnamefont
  {Motornenko}}, \bibinfo {author} {\bibfnamefont {A.}~\bibnamefont
  {Sorensen}}, \bibinfo {author} {\bibfnamefont {Y.}~\bibnamefont {Nara}},
  \bibinfo {author} {\bibfnamefont {V.}~\bibnamefont {Koch}}, \ and\ \bibinfo
  {author} {\bibfnamefont {M.}~\bibnamefont {Bleicher}},\ }\href@noop {} {\
  (\bibinfo {year} {2022})},\ \Eprint {http://arxiv.org/abs/2208.12091}
  {arXiv:2208.12091 [nucl-th]} \BibitemShut {NoStop}%
\bibitem [{\citenamefont {Sorge}\ \emph {et~al.}(1989)\citenamefont {Sorge},
  \citenamefont {Stoecker},\ and\ \citenamefont {Greiner}}]{Sorge:1989vt}%
  \BibitemOpen
  \bibfield  {author} {\bibinfo {author} {\bibfnamefont {H.}~\bibnamefont
  {Sorge}}, \bibinfo {author} {\bibfnamefont {H.}~\bibnamefont {Stoecker}}, \
  and\ \bibinfo {author} {\bibfnamefont {W.}~\bibnamefont {Greiner}},\ }\href
  {\doibase 10.1016/0375-9474(89)90641-6} {\bibfield  {journal} {\bibinfo
  {journal} {Nucl. Phys. A}\ }\textbf {\bibinfo {volume} {498}},\ \bibinfo
  {pages} {567C} (\bibinfo {year} {1989})}\BibitemShut {NoStop}%
\bibitem [{\citenamefont {Nara}\ \emph {et~al.}(2020)\citenamefont {Nara},
  \citenamefont {Maruyama},\ and\ \citenamefont {Stoecker}}]{Nara:2020ztb}%
  \BibitemOpen
  \bibfield  {author} {\bibinfo {author} {\bibfnamefont {Y.}~\bibnamefont
  {Nara}}, \bibinfo {author} {\bibfnamefont {T.}~\bibnamefont {Maruyama}}, \
  and\ \bibinfo {author} {\bibfnamefont {H.}~\bibnamefont {Stoecker}},\ }\href
  {\doibase 10.1103/PhysRevC.102.024913} {\bibfield  {journal} {\bibinfo
  {journal} {Phys. Rev. C}\ }\textbf {\bibinfo {volume} {102}},\ \bibinfo
  {pages} {024913} (\bibinfo {year} {2020})},\ \Eprint
  {http://arxiv.org/abs/2004.05550} {arXiv:2004.05550 [nucl-th]} \BibitemShut
  {NoStop}%
\bibitem [{\citenamefont {Steinheimer}\ \emph
  {et~al.}(2011{\natexlab{a}})\citenamefont {Steinheimer}, \citenamefont
  {Schramm},\ and\ \citenamefont {Stocker}}]{Steinheimer:2010ib}%
  \BibitemOpen
  \bibfield  {author} {\bibinfo {author} {\bibfnamefont {J.}~\bibnamefont
  {Steinheimer}}, \bibinfo {author} {\bibfnamefont {S.}~\bibnamefont
  {Schramm}}, \ and\ \bibinfo {author} {\bibfnamefont {H.}~\bibnamefont
  {Stocker}},\ }\href {\doibase 10.1088/0954-3899/38/3/035001} {\bibfield
  {journal} {\bibinfo  {journal} {J. Phys. G}\ }\textbf {\bibinfo {volume}
  {38}},\ \bibinfo {pages} {035001} (\bibinfo {year} {2011}{\natexlab{a}})},\
  \Eprint {http://arxiv.org/abs/1009.5239} {arXiv:1009.5239 [hep-ph]}
  \BibitemShut {NoStop}%
\bibitem [{\citenamefont {Steinheimer}\ \emph
  {et~al.}(2011{\natexlab{b}})\citenamefont {Steinheimer}, \citenamefont
  {Schramm},\ and\ \citenamefont {Stocker}}]{Steinheimer:2011ea}%
  \BibitemOpen
  \bibfield  {author} {\bibinfo {author} {\bibfnamefont {J.}~\bibnamefont
  {Steinheimer}}, \bibinfo {author} {\bibfnamefont {S.}~\bibnamefont
  {Schramm}}, \ and\ \bibinfo {author} {\bibfnamefont {H.}~\bibnamefont
  {Stocker}},\ }\href {\doibase 10.1103/PhysRevC.84.045208} {\bibfield
  {journal} {\bibinfo  {journal} {Phys. Rev. C}\ }\textbf {\bibinfo {volume}
  {84}},\ \bibinfo {pages} {045208} (\bibinfo {year} {2011}{\natexlab{b}})},\
  \Eprint {http://arxiv.org/abs/1108.2596} {arXiv:1108.2596 [hep-ph]}
  \BibitemShut {NoStop}%
\bibitem [{\citenamefont {Mukherjee}\ \emph {et~al.}(2017)\citenamefont
  {Mukherjee}, \citenamefont {Steinheimer},\ and\ \citenamefont
  {Schramm}}]{Mukherjee:2016nhb}%
  \BibitemOpen
  \bibfield  {author} {\bibinfo {author} {\bibfnamefont {A.}~\bibnamefont
  {Mukherjee}}, \bibinfo {author} {\bibfnamefont {J.}~\bibnamefont
  {Steinheimer}}, \ and\ \bibinfo {author} {\bibfnamefont {S.}~\bibnamefont
  {Schramm}},\ }\href {\doibase 10.1103/PhysRevC.96.025205} {\bibfield
  {journal} {\bibinfo  {journal} {Phys. Rev. C}\ }\textbf {\bibinfo {volume}
  {96}},\ \bibinfo {pages} {025205} (\bibinfo {year} {2017})},\ \Eprint
  {http://arxiv.org/abs/1611.10144} {arXiv:1611.10144 [nucl-th]} \BibitemShut
  {NoStop}%
\bibitem [{\citenamefont {Motornenko}\ \emph {et~al.}(2019)\citenamefont
  {Motornenko}, \citenamefont {Vovchenko}, \citenamefont {Steinheimer},
  \citenamefont {Schramm},\ and\ \citenamefont
  {Stoecker}}]{Motornenko:2018hjw}%
  \BibitemOpen
  \bibfield  {author} {\bibinfo {author} {\bibfnamefont {A.}~\bibnamefont
  {Motornenko}}, \bibinfo {author} {\bibfnamefont {V.}~\bibnamefont
  {Vovchenko}}, \bibinfo {author} {\bibfnamefont {J.}~\bibnamefont
  {Steinheimer}}, \bibinfo {author} {\bibfnamefont {S.}~\bibnamefont
  {Schramm}}, \ and\ \bibinfo {author} {\bibfnamefont {H.}~\bibnamefont
  {Stoecker}},\ }\href {\doibase 10.1016/j.nuclphysa.2018.11.028} {\bibfield
  {journal} {\bibinfo  {journal} {Nucl. Phys. A}\ }\textbf {\bibinfo {volume}
  {982}},\ \bibinfo {pages} {891} (\bibinfo {year} {2019})},\ \Eprint
  {http://arxiv.org/abs/1809.02000} {arXiv:1809.02000 [hep-ph]} \BibitemShut
  {NoStop}%
\bibitem [{\citenamefont {Motornenko}\ \emph {et~al.}(2020)\citenamefont
  {Motornenko}, \citenamefont {Steinheimer}, \citenamefont {Vovchenko},
  \citenamefont {Schramm},\ and\ \citenamefont
  {Stoecker}}]{Motornenko:2019arp}%
  \BibitemOpen
  \bibfield  {author} {\bibinfo {author} {\bibfnamefont {A.}~\bibnamefont
  {Motornenko}}, \bibinfo {author} {\bibfnamefont {J.}~\bibnamefont
  {Steinheimer}}, \bibinfo {author} {\bibfnamefont {V.}~\bibnamefont
  {Vovchenko}}, \bibinfo {author} {\bibfnamefont {S.}~\bibnamefont {Schramm}},
  \ and\ \bibinfo {author} {\bibfnamefont {H.}~\bibnamefont {Stoecker}},\
  }\href {\doibase 10.1103/PhysRevC.101.034904} {\bibfield  {journal} {\bibinfo
   {journal} {Phys. Rev. C}\ }\textbf {\bibinfo {volume} {101}},\ \bibinfo
  {pages} {034904} (\bibinfo {year} {2020})},\ \Eprint
  {http://arxiv.org/abs/1905.00866} {arXiv:1905.00866 [hep-ph]} \BibitemShut
  {NoStop}%
\bibitem [{\citenamefont {Li}\ \emph {et~al.}(2022)\citenamefont {Li},
  \citenamefont {Steinheimer}, \citenamefont {Reichert}, \citenamefont
  {Kittiratpattana}, \citenamefont {Bleicher},\ and\ \citenamefont
  {Li}}]{Li:2022iil}%
  \BibitemOpen
  \bibfield  {author} {\bibinfo {author} {\bibfnamefont {P.}~\bibnamefont
  {Li}}, \bibinfo {author} {\bibfnamefont {J.}~\bibnamefont {Steinheimer}},
  \bibinfo {author} {\bibfnamefont {T.}~\bibnamefont {Reichert}}, \bibinfo
  {author} {\bibfnamefont {A.}~\bibnamefont {Kittiratpattana}}, \bibinfo
  {author} {\bibfnamefont {M.}~\bibnamefont {Bleicher}}, \ and\ \bibinfo
  {author} {\bibfnamefont {Q.}~\bibnamefont {Li}},\ }\href@noop {} {\
  (\bibinfo {year} {2022})},\ \Eprint {http://arxiv.org/abs/2209.01413}
  {arXiv:2209.01413 [nucl-th]} \BibitemShut {NoStop}%
\bibitem [{\citenamefont {Huth}\ \emph {et~al.}(2022)\citenamefont {Huth} \emph
  {et~al.}}]{Huth:2021bsp}%
  \BibitemOpen
  \bibfield  {author} {\bibinfo {author} {\bibfnamefont {S.}~\bibnamefont
  {Huth}} \emph {et~al.},\ }\href {\doibase 10.1038/s41586-022-04750-w}
  {\bibfield  {journal} {\bibinfo  {journal} {Nature}\ }\textbf {\bibinfo
  {volume} {606}},\ \bibinfo {pages} {276} (\bibinfo {year} {2022})},\ \Eprint
  {http://arxiv.org/abs/2107.06229} {arXiv:2107.06229 [nucl-th]} \BibitemShut
  {NoStop}%
\bibitem [{\citenamefont {Randrup}(2004)}]{Randrup:2003mu}%
  \BibitemOpen
  \bibfield  {author} {\bibinfo {author} {\bibfnamefont {J.}~\bibnamefont
  {Randrup}},\ }\href {\doibase 10.1103/PhysRevLett.92.122301} {\bibfield
  {journal} {\bibinfo  {journal} {Phys. Rev. Lett.}\ }\textbf {\bibinfo
  {volume} {92}},\ \bibinfo {pages} {122301} (\bibinfo {year} {2004})},\
  \Eprint {http://arxiv.org/abs/hep-ph/0308271} {arXiv:hep-ph/0308271}
  \BibitemShut {NoStop}%
\bibitem [{\citenamefont {Vovchenko}\ \emph
  {et~al.}(2020{\natexlab{a}})\citenamefont {Vovchenko}, \citenamefont
  {Savchuk}, \citenamefont {Poberezhnyuk}, \citenamefont {Gorenstein},\ and\
  \citenamefont {Koch}}]{Vovchenko:2020tsr}%
  \BibitemOpen
  \bibfield  {author} {\bibinfo {author} {\bibfnamefont {V.}~\bibnamefont
  {Vovchenko}}, \bibinfo {author} {\bibfnamefont {O.}~\bibnamefont {Savchuk}},
  \bibinfo {author} {\bibfnamefont {R.~V.}\ \bibnamefont {Poberezhnyuk}},
  \bibinfo {author} {\bibfnamefont {M.~I.}\ \bibnamefont {Gorenstein}}, \ and\
  \bibinfo {author} {\bibfnamefont {V.}~\bibnamefont {Koch}},\ }\href {\doibase
  10.1016/j.physletb.2020.135868} {\bibfield  {journal} {\bibinfo  {journal}
  {Phys. Lett. B}\ }\textbf {\bibinfo {volume} {811}},\ \bibinfo {pages}
  {135868} (\bibinfo {year} {2020}{\natexlab{a}})},\ \Eprint
  {http://arxiv.org/abs/2003.13905} {arXiv:2003.13905 [hep-ph]} \BibitemShut
  {NoStop}%
\bibitem [{\citenamefont {Poberezhnyuk}\ \emph {et~al.}(2020)\citenamefont
  {Poberezhnyuk}, \citenamefont {Savchuk}, \citenamefont {Gorenstein},
  \citenamefont {Vovchenko}, \citenamefont {Taradiy}, \citenamefont {Begun},
  \citenamefont {Satarov}, \citenamefont {Steinheimer},\ and\ \citenamefont
  {Stoecker}}]{Poberezhnyuk:2020ayn}%
  \BibitemOpen
  \bibfield  {author} {\bibinfo {author} {\bibfnamefont {R.~V.}\ \bibnamefont
  {Poberezhnyuk}}, \bibinfo {author} {\bibfnamefont {O.}~\bibnamefont
  {Savchuk}}, \bibinfo {author} {\bibfnamefont {M.~I.}\ \bibnamefont
  {Gorenstein}}, \bibinfo {author} {\bibfnamefont {V.}~\bibnamefont
  {Vovchenko}}, \bibinfo {author} {\bibfnamefont {K.}~\bibnamefont {Taradiy}},
  \bibinfo {author} {\bibfnamefont {V.~V.}\ \bibnamefont {Begun}}, \bibinfo
  {author} {\bibfnamefont {L.}~\bibnamefont {Satarov}}, \bibinfo {author}
  {\bibfnamefont {J.}~\bibnamefont {Steinheimer}}, \ and\ \bibinfo {author}
  {\bibfnamefont {H.}~\bibnamefont {Stoecker}},\ }\href {\doibase
  10.1103/PhysRevC.102.024908} {\bibfield  {journal} {\bibinfo  {journal}
  {Phys. Rev. C}\ }\textbf {\bibinfo {volume} {102}},\ \bibinfo {pages}
  {024908} (\bibinfo {year} {2020})},\ \Eprint
  {http://arxiv.org/abs/2004.14358} {arXiv:2004.14358 [hep-ph]} \BibitemShut
  {NoStop}%
\bibitem [{\citenamefont {Vovchenko}\ \emph
  {et~al.}(2020{\natexlab{b}})\citenamefont {Vovchenko}, \citenamefont
  {Poberezhnyuk},\ and\ \citenamefont {Koch}}]{Vovchenko:2020gne}%
  \BibitemOpen
  \bibfield  {author} {\bibinfo {author} {\bibfnamefont {V.}~\bibnamefont
  {Vovchenko}}, \bibinfo {author} {\bibfnamefont {R.~V.}\ \bibnamefont
  {Poberezhnyuk}}, \ and\ \bibinfo {author} {\bibfnamefont {V.}~\bibnamefont
  {Koch}},\ }\href {\doibase 10.1007/JHEP10(2020)089} {\bibfield  {journal}
  {\bibinfo  {journal} {JHEP}\ }\textbf {\bibinfo {volume} {10}},\ \bibinfo
  {pages} {089} (\bibinfo {year} {2020}{\natexlab{b}})},\ \Eprint
  {http://arxiv.org/abs/2007.03850} {arXiv:2007.03850 [hep-ph]} \BibitemShut
  {NoStop}%
\bibitem [{\citenamefont {Bzdak}\ and\ \citenamefont
  {Koch}(2012)}]{Bzdak:2012ab}%
  \BibitemOpen
  \bibfield  {author} {\bibinfo {author} {\bibfnamefont {A.}~\bibnamefont
  {Bzdak}}\ and\ \bibinfo {author} {\bibfnamefont {V.}~\bibnamefont {Koch}},\
  }\href {\doibase 10.1103/PhysRevC.86.044904} {\bibfield  {journal} {\bibinfo
  {journal} {Phys. Rev. C}\ }\textbf {\bibinfo {volume} {86}},\ \bibinfo
  {pages} {044904} (\bibinfo {year} {2012})},\ \Eprint
  {http://arxiv.org/abs/1206.4286} {arXiv:1206.4286 [nucl-th]} \BibitemShut
  {NoStop}%
\bibitem [{\citenamefont {Savchuk}\ \emph {et~al.}(2020)\citenamefont
  {Savchuk}, \citenamefont {Poberezhnyuk}, \citenamefont {Vovchenko},\ and\
  \citenamefont {Gorenstein}}]{Savchuk:2019xfg}%
  \BibitemOpen
  \bibfield  {author} {\bibinfo {author} {\bibfnamefont {O.}~\bibnamefont
  {Savchuk}}, \bibinfo {author} {\bibfnamefont {R.~V.}\ \bibnamefont
  {Poberezhnyuk}}, \bibinfo {author} {\bibfnamefont {V.}~\bibnamefont
  {Vovchenko}}, \ and\ \bibinfo {author} {\bibfnamefont {M.~I.}\ \bibnamefont
  {Gorenstein}},\ }\href {\doibase 10.1103/PhysRevC.101.024917} {\bibfield
  {journal} {\bibinfo  {journal} {Phys. Rev. C}\ }\textbf {\bibinfo {volume}
  {101}},\ \bibinfo {pages} {024917} (\bibinfo {year} {2020})},\ \Eprint
  {http://arxiv.org/abs/1911.03426} {arXiv:1911.03426 [hep-ph]} \BibitemShut
  {NoStop}%
\bibitem [{\citenamefont {Sorensen}\ and\ \citenamefont
  {Koch}(2021)}]{Sorensen:2020ygf}%
  \BibitemOpen
  \bibfield  {author} {\bibinfo {author} {\bibfnamefont {A.}~\bibnamefont
  {Sorensen}}\ and\ \bibinfo {author} {\bibfnamefont {V.}~\bibnamefont
  {Koch}},\ }\href {\doibase 10.1103/PhysRevC.104.034904} {\bibfield  {journal}
  {\bibinfo  {journal} {Phys. Rev. C}\ }\textbf {\bibinfo {volume} {104}},\
  \bibinfo {pages} {034904} (\bibinfo {year} {2021})},\ \Eprint
  {http://arxiv.org/abs/2011.06635} {arXiv:2011.06635 [nucl-th]} \BibitemShut
  {NoStop}%
\bibitem [{\citenamefont {Kendall}\ and\ \citenamefont
  {Stuart}(1963)}]{kendall1963advanced}%
  \BibitemOpen
  \bibfield  {author} {\bibinfo {author} {\bibfnamefont {M.}~\bibnamefont
  {Kendall}}\ and\ \bibinfo {author} {\bibfnamefont {A.}~\bibnamefont
  {Stuart}},\ }\href {https://books.google.com/books?id=QhnvAAAAMAAJ} {\emph
  {\bibinfo {title} {The Advanced Theory of Statistics: Distribution
  theory}}},\ The Advanced Theory of Statistics\ (\bibinfo  {publisher} {Hafner
  Publishing Company},\ \bibinfo {year} {1963})\BibitemShut {NoStop}%
\bibitem [{\citenamefont {Vovchenko}(2021)}]{SMgithub}%
  \BibitemOpen
  \bibfield  {author} {\bibinfo {author} {\bibfnamefont {V.}~\bibnamefont
  {Vovchenko}},\ }\href@noop {} {\enquote {\bibinfo {title} {{A header-only C++
  library for mean and standard error estimation of moments and cumulants}},}\
  } (\bibinfo {year} {2021}),\ \bibinfo {note}
  {\href{https://github.com/vlvovch/sample-moments}{https://github.com/vlvovch/sample-moments}
  [Online; accessed 18-October-2022]}\BibitemShut {NoStop}%
\bibitem [{\citenamefont {Vovchenko}(2022)}]{Vovchenko:2022syc}%
  \BibitemOpen
  \bibfield  {author} {\bibinfo {author} {\bibfnamefont {V.}~\bibnamefont
  {Vovchenko}},\ }\href {\doibase 10.1103/PhysRevC.106.064906} {\bibfield
  {journal} {\bibinfo  {journal} {Phys. Rev. C}\ }\textbf {\bibinfo {volume}
  {106}},\ \bibinfo {pages} {064906} (\bibinfo {year} {2022})},\ \Eprint
  {http://arxiv.org/abs/2208.13693} {arXiv:2208.13693 [hep-ph]} \BibitemShut
  {NoStop}%
\bibitem [{\citenamefont {Adamczewski-Musch}\ \emph {et~al.}(2020)\citenamefont
  {Adamczewski-Musch} \emph {et~al.}}]{HADES:2020wpc}%
  \BibitemOpen
  \bibfield  {author} {\bibinfo {author} {\bibfnamefont {J.}~\bibnamefont
  {Adamczewski-Musch}} \emph {et~al.} (\bibinfo {collaboration} {HADES}),\
  }\href {\doibase 10.1103/PhysRevC.102.024914} {\bibfield  {journal} {\bibinfo
   {journal} {Phys. Rev. C}\ }\textbf {\bibinfo {volume} {102}},\ \bibinfo
  {pages} {024914} (\bibinfo {year} {2020})},\ \Eprint
  {http://arxiv.org/abs/2002.08701} {arXiv:2002.08701 [nucl-ex]} \BibitemShut
  {NoStop}%
\bibitem [{\citenamefont {Savchuk}\ \emph
  {et~al.}(2022{\natexlab{b}})\citenamefont {Savchuk}, \citenamefont
  {Poberezhnyuk},\ and\ \citenamefont {Gorenstein}}]{Savchuk:2022ljy}%
  \BibitemOpen
  \bibfield  {author} {\bibinfo {author} {\bibfnamefont {O.}~\bibnamefont
  {Savchuk}}, \bibinfo {author} {\bibfnamefont {R.~V.}\ \bibnamefont
  {Poberezhnyuk}}, \ and\ \bibinfo {author} {\bibfnamefont {M.~I.}\
  \bibnamefont {Gorenstein}},\ }\href {\doibase
  https://doi.org/10.1016/j.physletb.2022.137540} {\bibfield  {journal}
  {\bibinfo  {journal} {Physics Letters B}\ }\textbf {\bibinfo {volume}
  {835}},\ \bibinfo {pages} {137540} (\bibinfo {year}
  {2022}{\natexlab{b}})}\BibitemShut {NoStop}%
\bibitem [{\citenamefont {Bratkovskaya}\ \emph {et~al.}(2022)\citenamefont
  {Bratkovskaya}, \citenamefont {Glässel}, \citenamefont {Kireyeu},
  \citenamefont {Aichelin}, \citenamefont {Bleicher}, \citenamefont {Blume},
  \citenamefont {Coci}, \citenamefont {Kolesnikov}, \citenamefont
  {Steinheimer},\ and\ \citenamefont
  {Voronyuk}}]{https://doi.org/10.48550/arxiv.2208.11802}%
  \BibitemOpen
  \bibfield  {author} {\bibinfo {author} {\bibfnamefont {E.}~\bibnamefont
  {Bratkovskaya}}, \bibinfo {author} {\bibfnamefont {S.}~\bibnamefont
  {Glässel}}, \bibinfo {author} {\bibfnamefont {V.}~\bibnamefont {Kireyeu}},
  \bibinfo {author} {\bibfnamefont {J.}~\bibnamefont {Aichelin}}, \bibinfo
  {author} {\bibfnamefont {M.}~\bibnamefont {Bleicher}}, \bibinfo {author}
  {\bibfnamefont {C.}~\bibnamefont {Blume}}, \bibinfo {author} {\bibfnamefont
  {G.}~\bibnamefont {Coci}}, \bibinfo {author} {\bibfnamefont {V.}~\bibnamefont
  {Kolesnikov}}, \bibinfo {author} {\bibfnamefont {J.}~\bibnamefont
  {Steinheimer}}, \ and\ \bibinfo {author} {\bibfnamefont {V.}~\bibnamefont
  {Voronyuk}},\ }\href {\doibase 10.48550/ARXIV.2208.11802} {\enquote {\bibinfo
  {title} {Midrapidity cluster formation in heavy-ion collisions},}\ }
  (\bibinfo {year} {2022})\BibitemShut {NoStop}%
\bibitem [{\citenamefont {Kireyeu}\ \emph {et~al.}(2022)\citenamefont
  {Kireyeu}, \citenamefont {Steinheimer}, \citenamefont {Aichelin},
  \citenamefont {Bleicher},\ and\ \citenamefont
  {Bratkovskaya}}]{https://doi.org/10.48550/arxiv.2201.13374}%
  \BibitemOpen
  \bibfield  {author} {\bibinfo {author} {\bibfnamefont {V.}~\bibnamefont
  {Kireyeu}}, \bibinfo {author} {\bibfnamefont {J.}~\bibnamefont
  {Steinheimer}}, \bibinfo {author} {\bibfnamefont {J.}~\bibnamefont
  {Aichelin}}, \bibinfo {author} {\bibfnamefont {M.}~\bibnamefont {Bleicher}},
  \ and\ \bibinfo {author} {\bibfnamefont {E.}~\bibnamefont {Bratkovskaya}},\
  }\href {\doibase 10.48550/ARXIV.2201.13374} {\enquote {\bibinfo {title}
  {Deuteron production in ultra-relativistic heavy-ion collisions: A comparison
  of the coalescence and the minimum spanning tree procedure},}\ } (\bibinfo
  {year} {2022})\BibitemShut {NoStop}%
\end{thebibliography}%
\end{document}